\documentclass[aps,prb,superscriptaddress,twocolumn,showpacs,tightenlines,
amsmath,amsfonts]{revtex4}
\usepackage{graphicx,bm}
\usepackage{epstopdf}

\newcommand{\abs}[1]{\vert #1 \vert}

\newcommand{\M}{\mathbf{M}}
\newcommand{\m}{\mathbf m}
\renewcommand{\H}{\mathbf{H}}

\begin{document}

\title{Dynamics of a vortex domain wall in a magnetic nanostrip:
an application of the collective-coordinate approach}

\author{D. J. Clarke}
\affiliation{Department of Physics and
Astronomy, Johns Hopkins University, Baltimore, Maryland 21218, USA}
\author{O. A. Tretiakov}
\affiliation{Department of Physics and Astronomy, Johns Hopkins
University, Baltimore, Maryland 21218, USA}
\affiliation{Department of Physics, New York University, New York, New
York 10003, USA}
\author{G.-W. Chern}
\affiliation{Department of
Physics and Astronomy, Johns Hopkins University, Baltimore, Maryland
21218, USA}
\author{Ya. B. Bazaliy}
 \affiliation{Department of Physics and Astronomy, University of
South Carolina, Columbia, South Carolina 29208, USA}
 \affiliation{Institute of Magnetism, National Academy of Science of
Ukraine,
Kyiv 03142, Ukraine}
\author{O. Tchernyshyov}
\affiliation{Department of Physics and
Astronomy, Johns Hopkins University, Baltimore, Maryland 21218, USA}

\date{\today}

\begin{abstract}
The motion of a vortex domain wall in a ferromagnetic strip of
submicron width under the influence of an external magnetic field
exhibits three distinct dynamical regimes.  In a viscous regime at
low fields the wall moves rigidly with a velocity proportional to
the field.  Above a critical field the viscous motion breaks down
giving way to oscillations accompanied by a slow drift of the wall.
At still higher fields the drift velocity starts rising with the
field again but with a much lower mobility $dv/dH$ than in the
viscous regime.  To describe the dynamics of the wall we use the
method of collective coordinates that focuses on soft modes of the
system.  By retaining two soft modes, parametrized by the
coordinates of the vortex core, we obtain a simple description of
the wall dynamics at low and intermediate applied fields that
describes both the viscous and oscillatory regimes below and above
the breakdown.  The calculated dynamics agrees well with
micromagnetic simulations at low and intermediate values of the
driving field.  In higher fields, additional modes become soft and
the two-mode approximation is no longer sufficient. We explain some
of the significant features of vortex domain wall motion in high
fields through the inclusion of additional modes associated with the
half-antivortices on the strip edge.
\end{abstract}

\maketitle

\section{Introduction}

Dynamics of domain walls in ferromagnetic strips and rings with
submicron dimensions is a subject of active
research.\cite{Allwood05,Thomas07,Chien07} This topic is directly
relevant to several proposed schemes of magnetic memory and is also
interesting from the standpoint of basic physics. The dynamics of
domain walls under an applied magnetic field has distinct regimes:
viscous motion with a relatively high mobility at low fields and
underdamped oscillations with a slow drift at higher
fields.\cite{Beach05}

In a nanostrip, domain wall dynamics is further complicated by the
composite nature of the wall which consists of a few -- typically
two or three -- elementary topological defects in the bulk and at
the edge of the strip.\cite{OT05} As a result, its motion is
dominated by a few low-energy degrees of freedom associated with the
motion of the topological defects.  Weak external perturbations
engage only the softest (zero) mode -- a rigid translation of the
domain wall along the strip.  Larger external forces excite
additional modes thereby altering the character of motion.

The general approach to the dynamics of composite domain walls was
described recently by Tretiakov \textit{et al.}\cite{tretiakov:127204}
The configuration of a domain wall is parametrized by a few
collective coordinates $\bm \xi = \{\xi_1, \xi_2, \ldots, \xi_N\}$
representing the soft modes, and the free energy of the system $U$
is treated as a function of $\bm \xi$. The Landau-Lifshitz equation
for the spin dynamics with damping in Gilbert's form\cite{Gilbert55,Gilbert04}
is translated into a set of coupled equations of motion for the
collective coordinates. In a vector notation, they read
\begin{equation}
    \mathbf F - \Gamma \dot{\bm \xi} + G \dot{\bm \xi} = 0.
    \label{eq-main}
\end{equation}
Here components of the vector $\mathbf F$ are generalized forces
$F_i = -\partial U/\partial \xi_i$ derived from the free energy $U$;
the symmetric matrix $\Gamma_{ij}$ characterizes viscous friction;
the antisymmetric gyrotropic matrix $G_{ij}$ describes a
non-dissipative force of a kinematic origin related to spin
precession.

\begin{figure}
\includegraphics[width=\columnwidth]{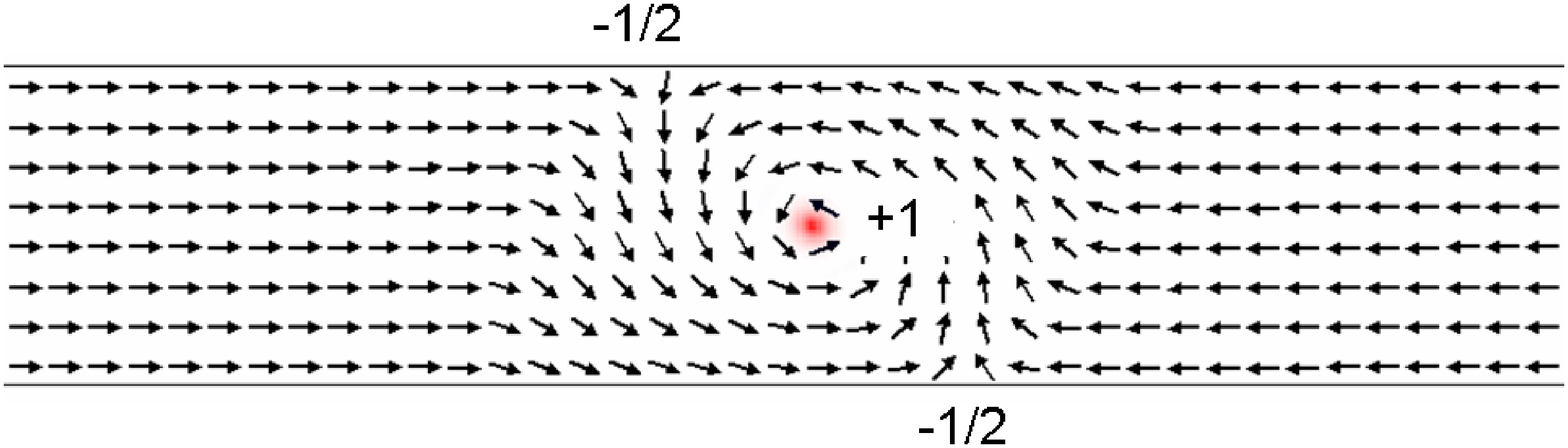}
\caption{(Color online) A head-to-head vortex domain wall in a
            long strip of permalloy 200 nm wide. The thickness in
            the out-of-plane direction is 20nm. Simulation using
            OOMMF.\cite{oommf} The winding numbers of the three
            topological defects are labeled. The shaded region
            indicates out-of-plane magnetization within the vortex
            core.}
\label{fig-oommfwall}
\end{figure}

In this paper we describe in detail the application of the
collective-coordinate method to the motion of a particular type of
domain walls, expanding on our previous
report.\cite{tretiakov:127204} Specifically, we focus on vortex
domain walls\cite{McMichael97} found in long permalloy strips of
submicron width $w$ and thickness $t \ll w$. In such a strip, the
magnetization pattern is essentially uniform across the thickness,
allowing us to consider patterns that vary only along the length
(x-direction) and width (y-direction) of the strip.

A vortex domain wall is a magnetization pattern that consists of
three topological defects (Fig.~\ref{fig-oommfwall}): a vortex in
the bulk of the strip and two half-antivortices confined to the
edges.\cite{Youk06} We will show that the seemingly complex dynamics
of a domain wall can be reduced to a simple motion of these defects.
The collective-coordinate approach focuses on the soft modes of a
system.  In the case of a vortex domain wall, the two softest modes
turn out to be the coordinate $(X)$ of the vortex core along the
long axis of the strip and the coordinate $(Y)$ of the core across
the strip width. The dynamic properties of the vortex are similar to
that of a massless charged particle moving through a viscous medium
in the presence of a potential $U(X,Y)$ and a fictitious magnetic
field directed normal to the plane. The ``electric charge" of the
particle equals $4\pi$ times the topological charge of the
vortex,\cite{tretiakov:127204} while the strength of the ``magnetic
field" equals the two-dimensional spin density.

At low fields the equations describe steady viscous motion of the
wall with a velocity proportional to the applied field. The vortex
is shifted in the transverse direction by an amount proportional to
the velocity of the wall.  At a certain critical field the vortex is
expelled from the strip and the steady motion breaks down, giving
way to an oscillatory regime where the vortex periodically crosses
the strip. Each time the vortex comes to an edge it is expelled and
reinjected to cross in the opposite direction. At very large
magnetic fields one reaches an extreme oscillatory regime, where the
vortex quickly moves back and forth across the strip almost along
the equipotential lines, while slowly drifting in the direction of
the applied field due to viscous forces.

The comparison of the predicted velocity curve to
experimental\cite{Beach05} and numerical\cite{tretiakov:127204} data
shows that the theory agrees well with the observations for low
fields and fields just above the breakdown. However, the theory
predicts a significantly lower drift velocity at high fields than is
observed. The origin of the discrepancy has been traced to the
appearance of additional soft modes in this regime that are not
taken into account in the two-mode approximation.  The new modes
increase the amount of dissipation in the wall and thereby lead to
faster drift.

The paper is organized as follows.  In Section~\ref{sec-cc} we review
the general aspects of the collective-coordinate approach in applications
to the dynamics of magnetization.
In Section~\ref{sec-2cc} we derive the equations of motion for
the two softest modes of the vortex wall parametrized by the
coordinates of the vortex core and discuss the general aspects of
their dynamics.  A detailed analysis of the dynamics is given in
Section~\ref{sec-wallmo}.  The additional soft modes are discussed in
Section \ref{sec-highmo}.  Auxiliary results are derived in the Appendixes.

\section{Collective Coordinates}\label{sec-cc}

\subsection{General formalism}

The dynamics of magnetization $\mathbf{M}(\mathbf r,t)$ in a
ferromagnet well below the Curie temperature is described by the
Landau-Lifshitz-Gilbert equation\cite{Gilbert55,Gilbert04} for the
unit vector $\m = \mathbf{M}/M$,
        \begin{equation}\label{eq-LLG}
            \frac{d \m}{d t}
            = \gamma \mu_0\H_{\mathrm{eff}} \times \m
            + \alpha \m \times \frac{d \m}{d t},
        \end{equation}
where $\mu_0\H_\mathrm{eff}(\mathbf r) = -\delta U/\delta
\mathbf{M(\mathbf r)}$ is an effective magnetic field. The
gyromagnetic ratio $\gamma = g|e|/(2m_e)$ is $1.75 \times 10^{11}
\mathrm{\ s \, A \, kg^{-1}}$ and the Gilbert damping constant
$\alpha \approx 0.01$ in permalloy.\cite{Freeman98}

The free energy includes, at the very least, exchange and dipolar
interactions as well as the Zeeman energy of the ferromagnet in an
external field $\mathbf{H}$,
    \begin{equation}
        U=\int dV \left(
        A\abs{\nabla\m}^2+\mu_0 H_{\rm{in}}^2/2-\mu_0 \H \cdot\M
        \right),
    \label{eq:U}
    \end{equation}
where $\mathbf{H}_{\rm{in}}$ is the field induced by the nonuniform
magnetization. It satisfies equations
$\nabla\cdot(\mathbf{H}_{\rm{in}} + \mathbf{M})=0$ and
$\nabla\times\mathbf{H}_{\rm{in}} =0$.  The exchange constant in
permalloy is $A=1.3\times10^{-13}$ J/m. Crystalline anisotropy is
negligibly small in permalloy; however, such terms can be included
in the free energy (\ref{eq:U}).

In principle, an infinite number of coordinates are necessary to
describe the time evolution of a magnetization texture. However, as
a domain wall propagates along a nanostrip, much of its structure
remains unchanged. While the wall may distort, it retains (at least
temporarily) its general shape. For instance, in a vortex wall the
chirality of the vortex does not change while the vortex remains
within the strip. Likewise, the magnetization far from the wall
remains fixed as the wall moves. This suggests that the motion of
the wall may be described by a finite set of coordinates $\bm \xi =
\{\xi_1, \xi_2, \ldots, \xi_N\}$, so that $\m = \m(\bm
\xi(t),\mathbf r)$.  In particular, the evolution of magnetization
is related to changes in the generalized coordinates as
    \begin{equation}\label{eq-cccond}
        \frac{d \mathbf{m}}{d t}=\frac{\partial \mathbf{m}}{\partial
        \xi_i}\dot{\xi_i},
    \end{equation}
where a sum over the repeated indices is implicit. By substituting
Eq.~(\ref{eq-cccond}) into Eq.~(\ref{eq-LLG}) and integrating over the
volume of the sample, we arrive at the equations of
motion\cite{tretiakov:127204}
    \begin{equation}\label{eq-cc}
        F_i-\Gamma_{ij}\dot{\xi}_j+G_{ij}\dot{\xi}_j=0,
    \end{equation}
with antisymmetric gyrotropic matrix $G_{ij}$ and symmetric matrix
of viscosity coefficients $\Gamma_{ij}$. Here
    \begin{subequations}
    \label{eq-ccdef}
    \begin{eqnarray}
    F_i&=& -\int dV \, \frac{\delta U}{\delta \m}\cdot\frac{\partial
                \m}{\partial
                \xi_i}=-\frac{\partial U}{\partial \xi_i},\\
    \Gamma_{ij}&=&\alpha J\int dV~
            \frac{\partial \m}{\partial \xi_i}\cdot\frac{\partial
                \m}{\partial \xi_j},\label{eq-ccdef2}\\
    G_{ij}&=&J\int dV~\m\cdot\frac{\partial \m}{\partial \xi_i}
    \times\frac{\partial \m}{\partial \xi_j},
    \end{eqnarray}
    \end{subequations}
and $J = M/\gamma$ is the density of angular momentum.

Note that $G_{ij}$ obeys the identity
    \begin{equation}
    \frac{\partial{G_{ij}}}{\partial{\xi_k}}
    +\frac{\partial{G_{jk}}}{\partial{\xi_i}}
    +\frac{\partial{G_{ki}}}{\partial{\xi_j}}=0.
    \end{equation}
If the space of collective coordinates is simply connected, then one
may express the gyrotropic tensor in terms of a gauge field $A_i$:
$G_{ij} = (\partial{A_j}/\partial{\xi_i}) -
(\partial{A_i}/\partial{\xi_j})$. The equations of motion
(\ref{eq-cc}) may then be derived from the Lagrangian
    \begin{equation}
        L=A_i\dot{\xi}_i-U,
    \label{eq:L}
    \end{equation}
together with the Rayleigh dissipation function
$R=\frac{1}{2}\dot{\xi}_i\Gamma_{ij}\dot{\xi}_j$.

\subsection{Soft and hard modes}\label{sec-soft}

Equations (\ref{eq-cc})--(\ref{eq-ccdef}) are formally exact when
they take into account all of the modes of a magnetic texture.  If
we are not interested in such level of detail, we may focus on a few
modes that capture the most salient features of magnetization
dynamics and neglect all other modes.  It is useful to divide modes
into soft ones, which remain active on the typical time scale $T$ of
the dynamics, and hard ones, the motion of which decays on a much
shorter time scale.  Since the drift velocity of a domain wall in an
applied magnetic field is ultimately determined by the rate at which
its Zeeman energy is dissipated, soft modes with long relaxation
times $\tau \gtrsim T$ are responsible for most of the dissipation
and thus control the drift velocity.  Steady-state motion has an
infinite characteristic time $T$, so that only the zero mode -- a
rigid translation of the wall -- with $\tau=\infty$ is relevant in
this case.  As shown in Sec.~\ref{sec-Xenergy}, the oscillatory
regime has a characteristic time scale $T = \pi/(\gamma \mu_0 H)$.
This regime has one additional soft mode with $\tau_1 \gtrsim T$ as
long as the field is not too strong. As the field strengthens, $T$
becomes shorter and eventually additional modes become soft, $\tau_2
\gtrsim T$ etc.

A large gyrotropic force creates a softening effect.  Consider, as
an illustration, system (\ref{eq-cc}) with two modes $\xi_{1,2}$,
free energy $U = k(\xi_1^2 + \xi_2^2)/2$, viscosity matrix
$\Gamma_{ij} = \Gamma \delta_{ij}$ and gyrotropic matrix $G_{ij} = G
\varepsilon_{ij}$, where $\varepsilon_{ij}$ is the antisymmetric
tensor with $\varepsilon_{12} = +1$. When $G = 0$, one has two
purely relaxational modes with $\tau = \Gamma/k$. These modes will
be soft for small stiffness $k$ or high viscosity $\Gamma$. In the
opposite limit $G \gg \Gamma$ the solution exhibits underdamped
oscillations with the relaxation time $\tau = G^2/(k \Gamma)$. The
latter exceeds the $G = 0$ result by a factor $(G/\Gamma)^2 \gg 1$.
We will see that in permalloy, the smallness of $\alpha\approx0.01$
means that $G \gg \Gamma$, i.e., the gyrotropic force indeed
dominates the viscous forces.

The general formalism described in this section is illustrated
in Appendix \ref{app:walker} on the classic problem of a Bloch
domain wall first considered by Walker\cite{Walker74} and recently
reviewed by Thiaville and Nakatani.\cite{ThiavilleBook06}

\section{Two-coordinate approximation: Generic features}\label{sec-2cc}

For domain walls under consideration, a large gyrotropic force is
associated with the motion of the vortex core (Appendix
\ref{app-gyro}). As a result, one may expect the vortex core motion
to represent the softest modes of the system, which is indeed the
case. We therefore consider a minimal description of the vortex wall
with just two coordinates $(X,Y)$ giving the location of the vortex
core.\cite{Youk06} In order to calculate, e.g., the viscosity tensor
$\Gamma_{ij}$, we must have a model ${\bf m} = {\bf m}(X(t),Y(t))$
for the wall. Such a model is discussed in
Appendix~\ref{app-models}, and the resulting viscosity components
for permalloy strips of widths $w=200$ and 600 nm are tabulated in
Table~\ref{table-Gamma}.
    \begin{table}
    \caption{Dimensionless ratios $\Gamma_{ij}/G$ for permalloy strips
    of widths $w=200$ and 600 nm computed for the vortex wall model of
    Youk \textit{et al.}\cite{Youk06} with vortex core placed in the
    center of the strip. Changing the position of the vortex does
    not change the results drastically. The ratios depend on the vortex
    chirality $\chi = \pm 1$.}
        \label{table-Gamma}
        \begin{tabular}{|l|c|c|c|}
            \hline
            & $\Gamma_{XX}/G$ & $\Gamma_{XY}/G$ & $\Gamma_{YY}/G$ \\
            \hline
            $w=200$ nm & 0.044 & $0.031\chi$ & 0.049\\
            \hline
            $w=600$ nm & 0.116 & $0.103\chi$ & 0.131\\
            \hline
        \end{tabular}
    \end{table}

However, once we have settled on the $X$ and $Y$ positions of the
vortex core as our collective coordinates, we may draw some general
conclusions regarding the motion that are independent of the model
we choose for the wall. In particular, this choice of coordinates
leads directly to a universal time for the vortex to cross the strip
in the transverse direction, in agreement with experimental
observations.\cite{hayashi:2006}

\subsection{The gyrotropic tensor}\label{sec-gyro}

    Because we are using only the two coordinates of the vortex core to
    describe the wall motion, the antisymmetric tensor $G_{ij}$ has a single
    independent component $G_{XY} = -G_{YX}$. As shown in
    Appendix~\ref{app-gyro}, as long as the vortex core is rigid, $G_{XY}$
    has a universal value $pG$, where $G=2\pi Jt$, $J = M/\gamma$ is the
    density of angular momentum, $t$ is the thickness of the film,
     and the polarization $p = \pm 1$ indicates
    the sign of the out-of-plane component of the magnetization within
    the vortex core.

\subsection{$X$-dependence of the free energy}
    \label{sec-Xenergy}

A vortex wall with the vortex core at $(X,Y)$ has the free energy $U
\approx U(X,Y)$. Because of the translational invariance along the
length of the strip, the dependence of the energy on the
longitudinal coordinate $X$ at a fixed $Y$ is trivial. A rigid shift
of the wall by $\Delta X$ alters the length of the two domains with
the opposite magnetizations by $+\Delta X$ and $-\Delta X$ and thus
changes the Zeeman energy $-\mu_0 \int \mathbf H \cdot \mathbf M \,
dV$ by $-QH \Delta X$, where $Q=2\mu_0 M t w$ is the magnetic charge
of the domain wall (see Appendix \ref{app-models}, Fig.~\ref{fig-gw}
in). Hence,
    \begin{equation}
        U(X,Y) = -QH X + U(Y).
    \end{equation}
Note that the longitudinal force $F_X = -\partial U/\partial X = QH$
is independent of the transverse coordinate $Y$ and in fact of any
other details of the wall structure.  This has interesting
consequences for wall motion in very high magnetic fields. In this
limit both gyrotropic and Zeeman forces are much larger than viscous
ones, i.e., such a regime corresponds to the limit of zero
viscosity. If viscous forces are completely neglected, the
longitudinal Zeeman force must balance the longitudinal component of
the gyrotropic force $pG\dot{Y}$ giving a constant transverse
velocity of the core $\dot Y = -pQH/G = -p\gamma\mu_0 H w/\pi$ and a
universal time\cite{JYLee07,arXiv:0711.3680v1} for the vortex to
cross the strip,
    \begin{equation}
        \label{eq-crosstime} T = \frac{w}{|\dot{Y}|} =
        \frac{\pi}{\gamma\mu_0 H}.
    \end{equation}
The transverse coordinate $Y$ thus oscillates at the Larmor
precession frequency $\omega_L = \gamma\mu_0 H$.  It is a remarkable
fact that in zero viscosity limit the same frequency is obtained for
a completely different domain wall in Walker's problem (see Appendix
\ref{app:walker} and Ref.~\onlinecite{Walker74}).

\subsection{$Y$-dependence of the free energy}\label{sec:Y-dependence}

In the absence of the applied field the system possesses a symmetry
with respect to 180$^{\circ}$-rotations of the strip, $X \mapsto -X$
and $Y \mapsto -Y$, so that $U(Y) \approx kY^2/2$ to the lowest
order in $Y$. (Reflectional symmetry $Y \mapsto -Y$ is broken by the
chiral nature of the vortex wall.) An applied magnetic field breaks
the rotational symmetry allowing a term linear in $Y$.  We thus have
    \begin{equation}\label{eq-U}
        U(X,Y) = -QH(X + r \chi Y) + kY^2/2 + \mathcal O(Y^3),
    \end{equation}
where $\chi = \pm 1$ is a chirality of the vortex and $r$ is a
numerical constant, and $k$ characterizes the restoring force that
tries to keep the vortex in the center of the strip. This results in
the transverse force $F_Y \approx r \chi QH - kY$. The model we use
for the vortex wall shows a significant transverse Zeeman force with
$r$ of order 1 (see Appendix~\ref{app-models}, Fig.~\ref{fig-gw}),
and nonzero restoring force.

\subsection{One-dimensional effective model}
    \label{sec-1Deff}

    In the absence of dissipation, the equations of motion may be derived
    from the Lagrangian
    \begin{equation}
        L(X,Y) = -pG\dot{X}Y-U(X,Y).
    \label{eq:L-2modes}
    \end{equation}
    Note that $\partial L/\partial \dot{X}=-pGY$, so $Y$ is the canonical
    momentum conjugate to $X$.

    If we are interested only in the longitudinal motion of the vortex, we may
    eliminate the transverse coordinate $Y$ using the equation of motion.
    In the harmonic approximation for the energy of the wall (\ref{eq-U}),
    $Y=(r \chi QH-pG\dot{X})/k$. Substituting this into Eq.~(\ref{eq:L-2modes})
    gives the effective Lagrangian representing a massive particle in one
    dimension subject to a constant external force,
    \begin{equation}
        L(X) = m\dot{X}^2/2 + QHX,
    \end{equation}
    with the effective mass $m = G^2/k$.  Using our estimate of the stiffness
    $k$, Eq.~(\ref{eq-k-compl}), and the definition of $G$ in
    Sec.~\ref{sec-gyro}, one finds that this mass is typically of order
    $10^{-22}$ kg.  This is the same order of magnitude as that found
    experimentally for a transverse wall by Saitoh \textit{et
    al.}\cite{Saitoh04} and theoretically for a one dimensional
    wall.\cite{Doring48, PhysRevLett.92.086601}
     The mass increases with the width of the strip and changes
    only weakly with the thickness.

    The effective one-dimensional description
    shall prove useful in Sec.~\ref{sec-highmo}, where additional
    degrees of freedom affect the motion of the wall. If these modes
    also have restoring forces acting on their conjugate momenta, the
    model becomes one of interacting massive particles in
    one dimension.

\section{Two-coordinate approximation: Dynamics}\label{sec-wallmo}

The equations of motion (\ref{eq-cc}) for two generalized coordinates
$\xi_0 = X$ and $\xi_1=Y$ read \cite{tretiakov:127204}
\begin{equation}\label{eom}
    F_i-\Gamma_{ij}\dot{\xi}_j + pG\varepsilon_{ij}\dot{\xi}_j = 0,
\end{equation}
where $\Gamma_{ij}=\Gamma_{ji}$ is the viscosity tensor, $p$ is the
polarization of the vortex core, and $\varepsilon_{ij}$ is the $2
\times 2$ antisymmetric tensor with $\varepsilon_{01}=+1$.  The
generalized forces $F_i=-\partial U/\partial \xi_i$ are derived from
the free energy (\ref{eq-U}). We thus arrive at the following
equations of motion for the vortex core:
\begin{subequations}
\label{eq-solvedeom}
\begin{eqnarray}
    \dot{X}&=&\frac{Q H}{\Gamma_{XX}}+\frac{k(\Gamma_{XY}- pG)}{G^2
    + \det\Gamma} \left(Y-Y_{\mathrm{eq}}\right),\\
    \dot{Y}&=&\frac{-k\Gamma_{XX}}{G^2 + \det\Gamma}
    \left(Y-Y_{\mathrm{eq}}\right),
\end{eqnarray}
\end{subequations}
where the equilibrium $Y$ position of the vortex is
\begin{equation}
    \label{eq}
    Y_\mathrm{eq} = \frac{(\chi g-p)GQH}{k\Gamma_{XX}}
\end{equation}
with
\begin{equation}\label{eq:g}
g = (r\Gamma_{XX}-\chi\Gamma_{XY})/G \ll 1 \ .
\end{equation}
It is worth noting that the magnitudes of the transverse
displacement $|Y_\mathrm{eq}|$ are slightly different for the two
possible values of the product $p\chi$ of the polarization and the
chirality. This effect can be traced to the lack of the reflection
symmetry $y \mapsto -y$ in a vortex wall, which leads to nonzero
transverse components of the Zeeman force $r\chi QH$ and the viscous
force $\Gamma_{YX}\dot{X}$. As a result, the trajectories of vortex
cores with the same polarization and opposite chiralities $\chi$ are
slightly different.

\begin{figure}
\includegraphics[width=\columnwidth]{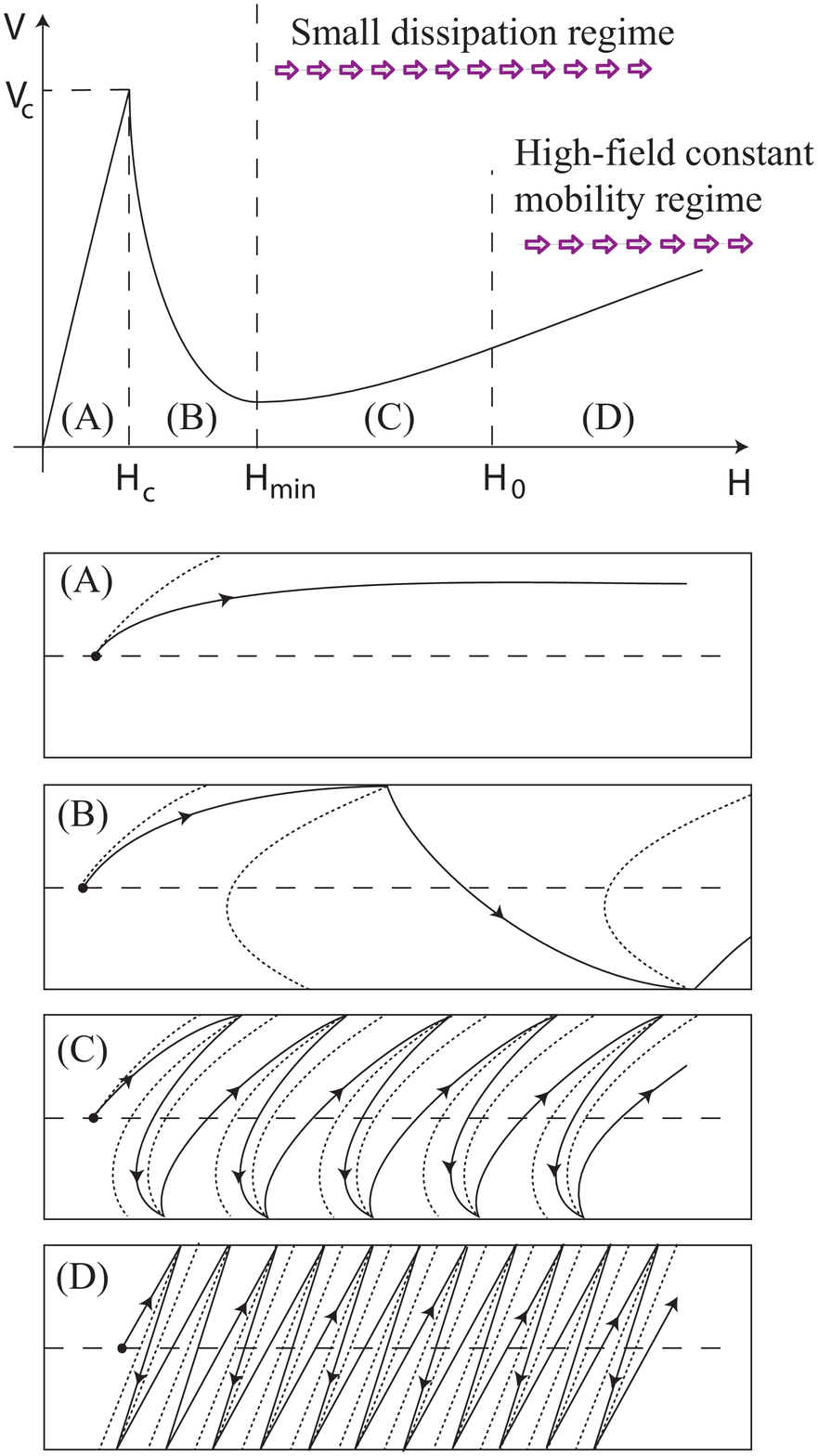}
\caption{A sketch of the domain wall drift velocity $V(H)$. Lower
panels show the vortex motion trajectory (solid lines) and
equipotential lines (dotted) at different field magnitudes. (A) -
below the Walker breakdown, (B) - just above the Walker breakdown,
(C) - small dissipation regime, (D) small dissipation regime with
dominating Zeeman force.}
 \label{fig-velocity_sketch}
\end{figure}

Analysis of equations of motion (\ref{eq-solvedeom}) yields three
distinct regimes (Fig.~\ref{fig-velocity_sketch}). Below a critical
field $H_c$ we find steady viscous motion with a high mobility $\mu
= dV/dH$ (see Sec.~\ref{sec-lowfield} below). Immediately above the
critical field $H_c$ the vortex motion becomes oscillatory, and the
drift velocity of the wall quickly decreases as the applied field
grows (see Sec.~\ref{full mobility} for details). At much higher
fields, $H \gg H_0 = k w/ 2 Q$, the drift velocity rises linearly
again but with a much lower mobility $\mu$ than at low fields
(Sections \ref{sec-genhighH} and \ref{sec-veryhighH} below). The
separation of scales $H_c$ and $H_0$ is guaranteed by the smallness
of parameter $\Gamma_{XX}/G$.

\subsection{Low fields ($H<H_c$)}\label{sec-lowfield}

    In a low applied field the wall exhibits simple viscous motion. The
    transverse coordinate of the vortex will asymptotically approach its
    equilibrium position $Y_\mathrm{eq}$, as long as the latter is within
    the strip.  The wall then moves rigidly with a steady longitudinal
    velocity $V = QH/\Gamma_{XX}$ giving the low-field mobility
    \begin{equation}\label{eq-lfm}
        \mu_\mathrm{LF} = d V/d H = Q/\Gamma_{XX}.
    \end{equation}
    Using the calculated value of $\Gamma_{XX}$ (\ref{eq-Gamma-youk})
    for a strip of width $w = 600$ nm and Gilbert damping $\alpha=0.01$
    yields $\mu_\mathrm{LF}= 29$ m s$^{-1}$ Oe$^{-1}$, which is not too far
    from the value of $25 \ \mathrm{m \, s^{-1} \, Oe^{-1}}$ measured
    by Beach \textit{et al.} \cite{Beach05}

\subsection{Critical field ($H=H_c$)}

    As demonstrated in Table~\ref{table-Gamma} and
    Eq.~(\ref{eq-Gamma-youk}), the viscous force is small in
    comparison with the gyrotropic one in permalloy strips with a
    width below 1 $\mu$m.  As a result, the equilibrium of a vortex
    in the transverse direction is set mostly by the balance of the
    transverse components of the gyrotropic force $GV$ and the
    restoring force $-kY_\mathrm{eq}$. The low-field regime ends when
    the equilibrium position of the vortex
    core is pushed beyond the strip edge, $|Y_\mathrm{eq}| \geq w/2$,
    making the steady state unreachable. The critical point is
    reached when $|Y_\mathrm{eq}| = w/2$, so that the critical velocity is
    \begin{equation}
        V_c \approx \frac{kw}{2G},
        \label{eq-crit}
    \end{equation}
    yielding the critical field
    \begin{equation}
        H_c = \frac{\Gamma_{XX}V_c}{Q} = \frac{kw \Gamma_{XX}}{2QG}.
    \label{eq-Hc}
    \end{equation}
    Beach \textit{et al.}\cite{Beach05} found a critical velocity of $80$ m/s in a
    permalloy wire $600$ nm wide. With the aid of Eq.~(\ref{eq-k-compl})
    for the stiffness constant $k$, we find $V_c=188$ m/s in such a wire.
    This is about twice as high as that observed experimentally. It is,
    however, much closer than that in Walker's one dimensional
    model of the wall, $1770$ m/s.\cite{Walker74,Yang08}  It is
    notable that our estimate lies between the experimental
    measurements and the critical velocity of 256 m/s,
    found in micromagnetic simulations by Yang \textit{et al.}\cite{Yang08}

    The critical velocity grows logarithmically with the width of the
    strip, and nearly linearly with its thickness. It is easy to see
    that the latter result is valid beyond the model of a vortex wall
    adopted in this calculation. The two forces balancing each other at
    $Y_\mathrm{eq} = w/2$ scale differently with $t$. While the
    gyrotropic force is linear in $t$, the restoring force comes from
    the magnetostatic energy, which represents Coulomb-like interaction
    of magnetic charges with density $\mathcal O(t)$, hence (the dipolar
    part of) the restoring force is quadratic in $t$ up to logarithmic
    factors.  This implies that $V_c$ is roughly proportional to $t$.

\subsection{General remarks for high fields
($H>H_c$)}\label{sec-genhighH}

Our numerical simulations in strips of thickness $t = 20$ nm and
width $w = 200$ nm indicate that, after the original vortex with a
core polarization $p$ is expelled from the strip, a new vortex is
injected at the same location with the opposite polarization $-p$.
The vortex thus moves between the edges switching its core
polarization each time it reaches an edge.

Once the transverse coordinate of the vortex $Y$ becomes a dynamical
variable, the motion acquires an entirely different character.  As
pointed out above, in permalloy strips the gyrotropic force $G
\mathbf V$ dwarfs the viscous force $\Gamma \mathbf V$. To zeroth
order in $\Gamma_{XX}/G$, the dynamics is purely conservative. Using
the general form of the Lagrangian (\ref{eq:L-2modes}) one can infer
that the vortex core moves along equipotential lines $U(X,Y) =
\mathrm{const}$. Absent viscosity, the vortex would oscillate back
and forth with the crossing time given by Eq.~(\ref{eq-crosstime})
but the wall would not move on average. Any total $x$-displacement
of the wall releases Zeeman energy, and thus requires viscous
friction to dissipate it.  At a small finite viscosity the vortex
trajectory slightly deviates from the equipotential lines, and the
wall slowly drifts in the longitudinal direction
(Fig.~\ref{fig-velocity_sketch} C,D).

\subsection{Very high fields ($H \gg H_0$)}\label{sec-veryhighH}

    We first demonstrate that at very high fields the velocity is again
    proportional to the field and calculate the high-field mobility for
    two-coordinate models. The new field scale $H_0$ is set by the
    requirement that the restoring force $-kY$ be negligible in comparison
    with the Zeeman force $QH$.  The characteristic field is
    \begin{equation}
        H_0 = \frac{kw}{2Q} = \frac{H_c G}{\Gamma_{XX}} \gg H_c.
    \end{equation}
When $H \gg H_0$, the dynamics is dominated by the Zeeman and
gyrotropic forces, so that the vortex moves almost along an
equipotential line $X+r\chi Y=\mathrm{const}$.

    As a result of the drift with a velocity $V_d$, the Zeeman
    energy goes down on average at the rate $QH V_d$.  It is
    dissipated through heat generated at the rate
    \[
        \mathbf V^T \Gamma \mathbf V = \dot{Y}^2
        \left(\begin{array}{cc}-r\chi & 1\end{array}\right) \hat{\Gamma}
        \left(\begin{array}{c}-r\chi \\ 1\end{array}\right).
    \]
    The transverse velocity of the vortex core reflects the balance
    between the longitudinal components of the gyrotropic and Zeeman
    forces: $\dot{Y} \approx -pQH/G$.  We thus find the drift velocity
    $V_d = \mu_\mathrm{HF}H$ with the high-field mobility
    \begin{equation}
        \mu_\mathrm{HF} = Q\left(\Gamma_{YY}
        -2r\chi\Gamma_{XY}+r^2\Gamma_{XX}\right)/G^2.
        \label{eq-hfm}
    \end{equation}

    The high-field (HF) mobility (\ref{eq-hfm}) is suppressed in
    comparison to the low-field (LF) one (\ref{eq-lfm}) by a factor of
    $\mathcal O(\Gamma^2/G^2) = \mathcal{O}(\alpha^2)$.  In the experiment
    of Beach
    \textit{et al.},\cite{Beach05} $\mu_\mathrm{HF}/\mu_\mathrm{LF}\approx
    0.1$. Using the $\Gamma$
    values for the more realistic model of Youk \textit{et al.}
    (shown in Table~\ref{table-Gamma}) and $r\approx2$ as predicted in the
    Appendix~\ref{app-models}, we find $\mu_\mathrm{HF}/\mu_\mathrm{LF}
    \approx 0.01$. Since the predicted low-field mobility matches the
    experimental value fairly well (see
    Sec.~\ref{sec-lowfield}), the calculated high-field mobility is much
    smaller than the observed value.

\subsection{High fields ($H>H_c$): Details}\label{full mobility}

To find the drift velocity of the vortex at fields above the vortex
expulsion field $H_{c}$, we determine the total $x$-displacement of
the vortex over a full cycle of motion from the top of the strip to
the bottom and back again. The crossing time will be slightly
different on the upward and downward trips due to a nonzero
transverse component of the Zeeman force
(Sec.~\ref{sec:Y-dependence}).

    Solving Eq.~(\ref{eom}) with polarization $p=\pm 1$ gives us the
    crossing times and displacements $\Delta T_+$ and $\Delta X_+$ (top to
    bottom) and $\Delta T_-$ and $\Delta X_-$ (bottom to top):
\begin{subequations}
\label{XandT}
\begin{eqnarray}
        \label{Xpm}
        \Delta X_\pm&=& \frac{QH \Delta T_\pm}{\Gamma_{XX}}
        -\frac{G \mp \Gamma_{XY}}{\Gamma_{XX}}w,
        \\
        \Delta T_\pm&=&2\frac{G^2 + \det\Gamma}{k\Gamma_{XX}}
        \mathrm{arctanh}\frac{H_{c\pm}}{H},
        \label{Tpm}
\end{eqnarray}
\end{subequations}
    where we define $H_{c\pm}=H_c/(1\mp\chi g)$, with $g$ defined by
    Eq.~(\ref{eq:g}). One can see that Eq.~(\ref{Tpm}) reduces to the
    universal crossing time of Eq.~(\ref{eq-crosstime}) in the limit
    $\Gamma_{ij}\to 0$. The drift velocity is
\begin{eqnarray}
V_d &=& \frac{\Delta X_+ +\Delta X_-}{\Delta T_+ + \Delta T_-}
\nonumber \\
&=& V_c \left(
        \frac{H}{H_c} - \frac{2(1+\det{\Gamma}/G^2)^{-1}}
        {\mathrm{arctanh}(H_{c+}/H)
        + \mathrm{arctanh} (H_{c-}/H)}\right).\nonumber \\
\label{eq-vd}
\end{eqnarray}
Note that the critical fields $H_{c\pm}$ are slightly different
reflecting a coupling between the vortex polarity $p$ and chirality
$\chi$ seen in Eq.~(\ref{eq}).

An expansion of Eq.~(\ref{eq-vd}) in powers of $1/H$ yields the
high-field result (\ref{eq-hfm}). The same expansion allows to find
the field $H_{\min}$ at which the domain wall velocity has a
minimum:
\begin{equation}\label{eq:Hmin}
\frac{H_{\min}}{H_c} \approx \frac{1}{\sqrt{3 (g^2 +
\det\Gamma/G^2)}} \sim \sqrt\frac{G}{\Gamma} \ .
\end{equation}
Using expressions (\ref{Xpm}) for the displacements $\Delta X_{\pm}$
it is possible to characterize the fields at which the vortex
trajectory approaches the zero-dissipation limit. Typical
trajectories of the vortex are shown in
Fig.~\ref{fig-velocity_sketch}(C,D). They are close to the
equipotential lines $U(X,Y) = {\rm const}$ when the displacement
$\Delta X_{+} + \Delta X_{-}$ in one back and forth cycle is
negligible compared to the strip width $w$. Expansion in $1/H$ gives
a criteria
$$
\frac{2G}{\Gamma_{XX}}\left(g^2 + \det\Gamma/G^2\right) +
\frac{2G}{3\Gamma_{XX}} \left(\frac{H_c}{H}\right)^2 \ll 1
$$
or
\begin{subequations}
\label{eq:zero_dissipation_criteria}
\begin{eqnarray}
&& \frac{2G}{\Gamma_{XX}}\left(g^2 + \det\Gamma/G^2\right) \ll 1
 \\
 \label{eq:zd1}
&& \frac{H}{H_c} \gg \sqrt{\frac{2G}{3\Gamma_{XX}}}
\end{eqnarray}
\end{subequations}
For $r \lesssim 1$ the first inequality is always satisfied when
$\Gamma \ll G$ holds. Eqs.~(\ref{eq:Hmin}) and (\ref{eq:zd1}) show
that vortex motion becomes approximately dissipationless for fields
above the velocity minimum.

\begin{figure}
\includegraphics[width=0.99\columnwidth]{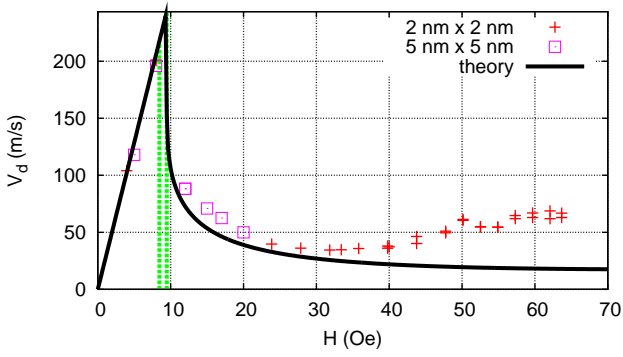}
\caption{(Color online) The drift velocity $V_d$ of the domain wall
            as a function of the applied field $H$ for a permalloy
            strip of width $w=200$ nm and thickness $t=20$ nm.
            Dashed vertical lines mark the critical fields $H_{c-}$
            and $H_{c+}$.}
\label{fig-numerics}
\end{figure}

In Fig.~\ref{fig-numerics} the predicted drift velocity is compared
with the results of numerical simulations for a permalloy strip of
width $w=200$ nm and thickness $t=20$ nm.\cite{tretiakov:127204}

    The components of the viscosity tensor used are the predicted values
    for the Youk \textit{et al.} model listed in Table~\ref{table-Gamma}.
    \begin{figure}[htbp]
    \includegraphics[width=\columnwidth]{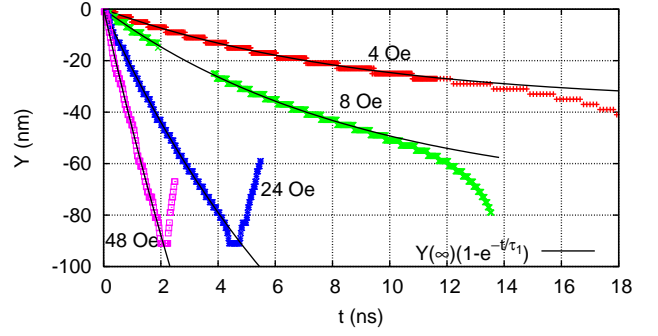}
\caption{(Color online) The time constant $\tau_1$ is determined by
fitting the numerically calculated time dependence of the
$Y$-coordinate to a decaying exponential, see Eq.~(\ref{eq-exp}).}
    \label{fig-ynum}
    \end{figure}
    The stiffness constant $k$ of the restoring potential could not be
    calculated accurately because two of its main contributions, a
    positive magnetostatic term and a negative term due to N\'eel-wall
    tension, are of the same order of magnitude. This is not surprising
    given the proximity of the strip used in the simulations to a region
    where the vortex wall is unstable.\cite{McMichael97} Instead, the
    relaxation time $\tau_1=(G^2 + \det\Gamma)/k\Gamma_{XX}$ was extracted
    directly from the numerics by fitting $Y(t)$ to the exponential time
    dependence described by Eq.~(\ref{eq-solvedeom}), i.e.
    \begin{equation}
        Y(t)=Y_\mathrm{eq}+[Y(0)-Y_\mathrm{eq}]\exp(-t/\tau_1),
        \label{eq-exp}
    \end{equation}
    see Fig.~\ref{fig-ynum}.  For applied fields from 4 to 60 Oe, $\tau_1$
    was in the range from 8.5 to 9 ns. This leads to a $k$ value roughly twice
    that predicted by Eq.~(\ref{eq-k-compl}). In accordance with
    Eq.~(\ref{eq-solvedeom}), the equilibrium position $Y_{\mathrm{eq}}$
    scales linearly with $H$. In calculating the critical velocity
    $V_c=kw/(2G)$, it was necessary to replace $w$ with an effective strip
    width $w_\mathrm{eff} = w-2R_0$ to account for a finite size of a
    vortex core\cite{Wachowiak02} and edge defect. As the vortex
    approaches the edge, the short-range attractive exchange interaction
    overwhelms the dipolar potential and draws the vortex into the edge
    defect. This leads to a deviation of the transverse vortex coordinate
    plotted in Fig.~\ref{fig-ynum} from Eq.~(\ref{eq-exp}) as the vortex
    comes close to the edge. From the observed trajectories we estimated
    $R_0 \approx 10$ nm.

\section{Beyond the two-coordinate approximation}\label{sec-highmo}

\begin{figure}
\includegraphics[width=\columnwidth]{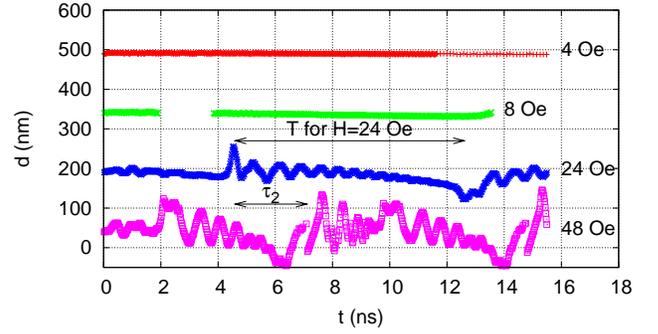}
\caption{(Color online) Numerical simulations above the critical
field show repeated oscillations in the width of the wall, measured
as the distance between the half-antivortices. The patterns for
various field values are offset vertically by 150 nm for clarity.
Each wall width is oscillating near $w=200$ nm. The vortex crossing
time $T$ and the decay time of wall width oscillations $\tau_2$ are
shown by arrows.}
        \label{fig-width-osc}
\end{figure}

At low and intermediate fields, the two-coordinate model shows good
agreement with numerical simulations.  However, in higher fields the
calculated drift velocity falls below the value observed in the
numerics (Fig.~\ref{fig-numerics}).  The discrepancy is caused by
the softening of additional degrees of freedom that have so far been
ignored.  The new soft modes provide additional channels for the
dissipation of Zeeman energy leading to an increase in the velocity.
In fields above $H_2 = 35$ Oe, the oscillation period $T =
\pi/(\gamma\mu_0 H)$ becomes comparable to the decay times $\tau$ of
some of the harder modes. The new modes can be seen in oscillations
of the total width of the wall that occur after the vortex has been
absorbed by the edge and reemitted (Fig.~\ref{fig-width-osc}).
Including additional modes in the calculation clarifies some of the
significant features of the wall dynamics observed in numerical
simulations at high fields.

We first present a hypothetical toy model with two additional
degrees of freedom in Sec.~\ref{sec-hardmodes}.  It provides a
pedagogical example of coupled hard and soft modes with gyrotropic
forces and sets the stage for a more realistic model with six
degrees of freedom appropriate for a vortex domain wall in higher
fields, which is the subject of Sec.~\ref{sec-3mode}.

\subsection{Pedagogical model with 4 coordinates}
\label{sec-hardmodes}

    The high-frequency oscillations of a vortex wall (Fig.~\ref{fig-width-osc})
    can be reproduced in a simple phenomenological model where the wall
    is treated as a vortex-antivortex pair with cores located at $(x_1, y_1)$
    and $(x_2, y_2)$.  (A real vortex wall contains a vortex and two
    half-antivortices.\cite{OT05})  The gyrotropic tensor of the system has
    the following nonzero components:
    \begin{equation}
        G_{x_1 y_1} = -G_{y_1 x_1} = G_1,
        \quad G_{x_2 y_2} = -G_{y_2 x_2} = G_2.
    \end{equation}
    Both defects are confined in the transverse direction and coupled to
    each other in the longitudinal direction with an equilibrium
    separation $a$.  They carry magnetic charges $Q_1$ and $Q_2$.  The
    potential energy of the system is
    \begin{equation}
        U = \sum_{i=1}^2 \left(\frac{k_i y_i^2}{2} - H Q_i x_i\right)
        + \frac{k (x_1-x_2-a)^2}{2}. \label{eq-U-2m}
    \end{equation}
    The Lagrangian is
    \begin{equation}\label{eq-L-2m}
        L= -G_1 \dot{x}_1 y_1-G_2 \dot{x}_2 y_2-U.
    \end{equation}
    In this pedagogical example we assume that the viscous force acts on the
    antivortex only, so that the dissipation tensor has a single nonzero
    component $\Gamma_{x_2 x_2} = \Gamma > 0$, giving rise to the Rayleigh
    dissipation function $R = \Gamma \dot{x}_2^2/2$.  Also, it is
    important to note that the vortex core is flipped whenever it reaches
    the edge of the strip, $y_1 = \pm w/2$, altering the sign of $G_1$.

    \subsubsection{Two massive particles}
    \label{sec-2mass}

        It is instructive to eliminate the transverse coordinates $y_1$ and
        $y_2$, as in Sec.~\ref{sec-1Deff}, using the equations of motion, $k_i
        y_i=-G_i\dot{x}_i$.  The resulting dynamics is that of two massive
        particles moving in one dimension. The potential energy of the
        original model (\ref{eq-U-2m}) translates into a sum of kinetic energy
        $K$ and potential energy $U$ of the massive particles:
        \begin{eqnarray}
        E &=& K+U
        \nonumber \\
        &=& \frac{m_1 \dot{x}_1^2}{2} + \frac{m_2 \dot{x}_2^2}{2}
        + \frac{k (x_1 - x_2 - a)^2}{2} -H(Q_1 x_1 + Q_2 x_2), \nonumber \\
        \label{eq-E}
        \end{eqnarray}
        where the masses are determined by the gyrotropic and
        stiffness constants, $m_i = G_i^2/k_i$. Continuous evolution
        of the system is disrupted when the velocity of the vortex
        attains a critical magnitude, $|\dot x_1| = v_c = k_1
        w/2G_1$, signaling that the vortex has reached the edge. At
        that moment $G_1$ changes sign and the longitudinal vortex
        velocity $\dot{x}_1 = -k_1 y_1/G_1$ is reversed, $\dot x_1
        \to -\dot x_1$. After that continuous evolution resumes.
        Nothing special happens to the antivortex at that moment.

        The natural modes of the system are the center of mass $X =
        (m_1 x_1 + m_2 x_2)/M$ and relative position $x = x_1 - x_2
        - a$.  Their energies are
        \begin{subequations}
        \label{eq-E2}
        \begin{eqnarray}
                    E_\mathrm{CM} &=& \frac{M \dot{X}^2}{2} - QHX,
                    \\
                    E_\mathrm{rel} &=& \frac{\mu \dot{x}^2}{2}
                    + \frac{kx^2}{2} - qHx,
        \end{eqnarray}
        \end{subequations}
        where $M = m_1 + m_2$ and $\mu = (m_1^{-1} + m_2^{-1})^{-1}$ are the
        total and reduced masses and $Q = Q_1 + Q_2$ is the total magnetic
        charge. The value of the relative magnetic charge $q$ will not matter
        to our analysis.

        We consider the limit in which the antivortex is strongly
        confined in the transverse direction, so that $k_2 \gg k_1$
        and as a result, the antivortex is much lighter than the
        vortex,
        \begin{equation}
            M \sim m_1 \gg m_2 \sim \mu.
        \end{equation}
        Here we rely on the fact that the gyrotropic coefficients of
        the vortex and the antivortex have the same magnitude, see
        Appendix \ref{app-gyro}.

        The center-of-mass position $X$ is a zero mode with an
        infinite relaxation time.  Its velocity $\dot{X}$ is an
        overdamped mode with the relaxation time $\tau_1 \sim
        m_1/\Gamma = G_1^2/(k_1 \Gamma)$, which is essentially the
        same as in our two-coordinate model of the vortex wall. The
        remaining two modes represent underdamped oscillations of
        the relative coordinate and velocity with the frequency
        $\omega \sim \sqrt{k/\mu}$ and relaxation time $\tau_2\sim
        2m_2/\Gamma \ll \tau_1$.  Thus $x$ and $\dot x$ are much
        harder than $X$ and $\dot X$ and there is a regime, $\tau_1
        \gg T \gg \tau_2$, where we may neglect the relative motion
        as a hard mode.  In that regime, the dynamics of the wall
        reduces to that of our simple two-mode model described in
        Sec.~\ref{sec-wallmo}.

\subsubsection{Losses from the hard modes}

        Let us compare the energy losses from the soft and hard
        modes to check whether our neglect of the hard modes is
        justified.  In the limit $\tau_2 \ll T$, the relative
        position $x$ and velocity $\dot x$ quickly reach their
        equilibrium values retaining them through most of the
        crossing period $T$.  However, as the velocity of the vortex
        $\dot{x}_1$ reverses its sign, the relative velocity
        $\dot{x} = \dot{x}_1-\dot{x}_2$ changes as well increasing
        from zero to $\pm 2v_c$.  The energy of relative motion
        increases by $2 \mu v_c^2$ at the expense of the
        center-of-mass energy.
        (Note that the total energy (\ref{eq-E}) does not change at
        all since it is not sensitive to the sign of $\dot{x}_1$.)
        The kinetic energy of relative motion $2 \mu v_c^2$ will be fully
        dissipated during the initial stage of the next crossing.

        The energy lost by the oscillatory relative motion, $\Delta
        E_2 = 2 \mu v_c^2$, should be compared to the total loss
        incurred primarily through the overdamped main mode,
        \begin{equation}
            \Delta E = \int_0^T  \Gamma \dot{x}_2^2 dt
            \sim \int_0^{T} \Gamma \dot{X}^2 dt
            \sim \frac{\Gamma v_c^2 T}{3}.
        \end{equation}
        The fraction of energy dissipated through the hard modes is
        thus
        \begin{equation}
            \label{eq-relE}
            \frac{\Delta E_2}{\Delta E} \sim \frac{3\tau_2}{T}.
        \end{equation}
        As long as relative motion remains hard, $\tau_2 \ll T$,
        energy loss associated with it is negligible and the motion
        of the system is well approximated by the soft modes $X$ and
        $\dot X$ alone.

\subsection{Model with 6 coordinates}
\label{sec-3mode}

Transient oscillations of the hard modes found in the pedagogical
model are similar to the oscillations of the width of the vortex
wall (Fig.~\ref{fig-width-osc}), measured as the distance between
the two edge defects. In our numerical simulations, after the vortex
is absorbed and reemitted at the edge, the width of the wall
displays underdamped oscillations.  In moderate fields, these
oscillations decay during the crossing time $T$ and are reactivated
at the beginning of the next cycle.  This activation is evidence of
energy transfer from the vortex core to harder modes, just as
occurred in the pedagogical model of Sec.~\ref{sec-hardmodes}. In
sufficiently high fields, the vortex crossing time $T =
\pi/(\gamma\mu_0 H)$ becomes comparable to the relaxation time of
the new modes. Notably, this occurs in the same region of field
strength in which the two-coordinate model prediction for the drift
velocity begins to deviate from the results of numerical simulations
(Fig.~\ref{fig-numerics}). New dissipation channels become important
in higher fields, changing the dynamics of the wall.

    \begin{figure}
    \includegraphics[width=\columnwidth]{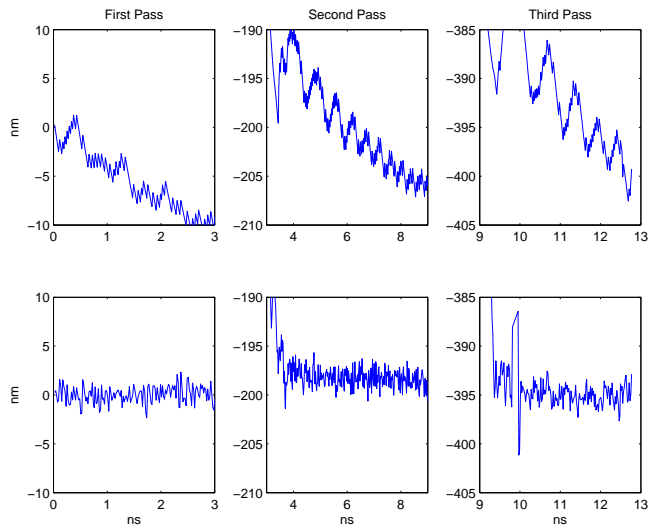}
    \caption{(Color online) Top row: The left hand side of
        Eq.~(\ref{eq-xmom2}) for the two-coordinate model plotted
        for the first three passes of the vortex across the strip
        under an applied field of 32 Oe.   Bottom Row: When a finite
        mass $m = 1.1\times10^{-23}$ kg is included for the edge
        defects, as in Eq.~(\ref{eq-xmom4}), the total $x$-momentum
        of the modes is conserved.} \label{fig-edgemass}
    \end{figure}

    Further evidence of the need to include additional modes
    may be gleaned by comparing the equations of motion of the
    two-coordinate model to the numerical simulations directly.
    The two-coordinate model predicts that
    \begin{equation}\label{eq:x-momentum}
        -pG\dot{Y}=QH-\Gamma_{XX}\dot{X}.
    \end{equation}
This equation expresses the change of linear momentum in the $x$
direction due to external forces, and holds while the vortex is
inside the strip. Indeed, the momentum canonically conjugate to
coordinate $X$ is $\partial L/\partial \dot{X}=-pGY$
(\ref{eq:L-2modes}). As a consequence of the translational
invariance of the strip in the $x$ direction, this momentum is
influenced only by forces external to the wall, i.e. the driving
force $QH$ from the field and the drag force $-\Gamma_{XX}\dot{X}$.
Equation (\ref{eq:x-momentum}) may be integrated to yield
    \begin{equation}
        \frac{\Gamma_{XX}}{G}X-pY-\frac{QH}{G}t =\mathrm{const}
        \label{eq-xmom2}
    \end{equation}
The above holds while the vortex is crossing the strip. Formally
Eq.~(\ref{eq-xmom2}) does not describe the collision with the edge,
but it can be checked that the effect of collision and corresponding
vortex polarization flip is captured by a change of constant on the
right hand side of Eq.~(\ref{eq-xmom2}) by $\mathrm{const} \to
\mathrm{const} \pm w$.

    We can test this prediction using numerical simulations. The
    value of the left-hand-side of this equation is plotted in the
    top row of Fig.~\ref{fig-edgemass} for the first three passes of
    the vortex across a 200-nm-wide, 20-nm-thick strip when a
    32-Oe driving field is applied. For $\Gamma_{XX}$, we use the
    theoretical estimate $\Gamma_{XX}=0.044G$ for the model of Youk
    \textit{et al.} (Table~\ref{table-Gamma}). In the two-mode
    approximation, the plotted value should be constant during each
    pass. However, it is clearly not constant and displays
    oscillations reminiscent of those in the wall width
    (Fig.~\ref{fig-width-osc}).

Equation~(\ref{eq-xmom2}), derived in the two-mode approximation
states that momentum along the nanostrip is influenced only by two
external forces. Its violation reflects a transfer of momentum
between the vortex and other modes of the wall. In a model with a
larger number of modes a similar equation on the x-direction
momentum can be derived and checked numerically. It provides a good
test showing whether the new set of modes is sufficient to capture
the actual wall motion.

    The evidence in Fig.~\ref{fig-width-osc} and
    Fig.~\ref{fig-edgemass} shows that the modes associated with the
    positions of the edge defects become important to the motion of
    the wall in intermediate to high fields. We add the coordinates of
    these defects to our description in order to explain qualitatively
    their behavior as observed in numerical simulations. We use some
    simple modeling for the forces holding the edge defects in place
    relative to the vortex core. As with the vortex itself, each edge
    defect is described by two coordinates that are coupled to one
    another through the gyrotropic tensor. In the case of the vortex,
    these two variables are the $X$ and $Y$ coordinates of the
    core. For the edge defects, we use the $X$-positions $X_\pm$ of
    the defects and the out-of-plane angles $\theta_\pm$ of the
    magnetization at their cores. It is not surprising that a
    gyrotropic coupling should be present between these coordinates,
    as it is the same pairing that occurs in Walker's problem of the
    Bloch wall---see Appendix A for details.

    \subsubsection{Energy}

        In order to describe the motion of the two edge defects, two
        springs are added to the energy $U(X,Y)$ in Eq.~(\ref{eq-U})
        that act to keep the upper $(X_+)$ and lower $(X_-)$ edge
        defects on a $45^\circ$ line
        with the vortex
        \begin{equation}\label{eq-springs}
            U_{e\pm}=\frac{k_2}{4}
            \left(X+\chi Y-X_\pm \mp\frac{\chi w}{2}\right)^2.
        \end{equation}
        Here again $\chi$ is the chirality of the vortex.  (Which
        of the two edge defects is ahead of the vortex depends on
        the vortex chirality.)

        To take into account the energy cost of bringing magnetization
        out of the plane of the strip near the edge defects, the
        restoring springs are added for the angles $\theta_+$ and
        $\theta_-$ associated with the upper and lower edge defects,
        respectively:
        \begin{equation}
            U_{\theta\pm}=\frac{k_\theta}{2}\theta_\pm^2.
        \end{equation}

        \begin{figure}[htbp]
        \includegraphics[width=\columnwidth]{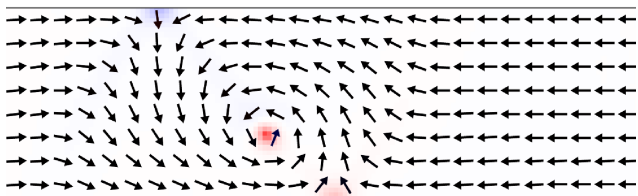}
        \caption{(Color online) Image of a wall moving to the right
                 with an applied field of 4 Oe. Shaded regions
                 indicate out-of-plane magnetization.}
        \label{fig-tilts}
        \end{figure}
        To represent the gyrotropic coupling of the positions of the
        edge defects to their out-of-plane angles, the terms
        \begin{equation}\label{Llong}
            L_\pm = \mp G_2\dot{X}_\pm\theta_\pm
        \end{equation}
        are added to the Lagrangian.

        The gyrotropic coefficients for the upper and lower edge
        defects have opposite signs because they wind in opposite
        directions as one moves from left to right. This effect
        is visible in simulations even below the first breakdown
        field (Fig.~\ref{fig-tilts}). As the wall moves to the
        right at a steady velocity, the lower defect tilts
        up out of the plane, while the upper defect tilts into the
        plane.

        The total Lagrangian for this extended model is then
        \begin{equation}\label{eq-Lex}
            L = -pG\dot{X}Y -U(X,Y)
            +\sum_{b=\pm} (L_b-U_{e b}-U_{\theta b}),
        \end{equation}
        where $U(X,Y)$ is given by Eq.~(\ref{eq-U}).

    \subsubsection{Viscosity}

        It remains to determine the components of the viscosity
        tensor for the three defects.  Under the assumption that the
        three defects may move independently of one another, it is
        logical to assume that the
        viscosity tensor $\Gamma$ will be essentially diagonal. Each
        of the edge defects carries with it a portion of the wall,
        and that portion's associated viscosity. Since the vortex is
        moving independently of the edge defects, there is no
        asymmetry that would lead to off-diagonal terms.  The most
        general (second order) Rayleigh dissipation function
        consistent with these conditions is:
        \begin{equation}\label{eq-R}
            R=\frac{1}{2}\sum_i\Gamma_i\dot{\xi_i}^2,
        \end{equation}
where $\Gamma_{X_+}=\Gamma_{X_-}=\Gamma_e$, and
$\Gamma_{\theta_+}=\Gamma_{\theta_-}=\Gamma_\theta$.  The largest
dissipation component is expected to come from the $X_{\pm}$
positions of the half-antivortices, as they influence the motion of
the N\'eel walls that emanate from them (see Fig.~\ref{fig-youk}).
These N\'eel walls carry most of the viscosity of the wall, as they
are the regions where the magnetization changes most rapidly with
position. According to Eq.~(\ref{eq-ccdef2}), the viscous tensor
components have the largest contribution from the regions where the
change of magnetization with the coordinate is the largest.  The
smallest dissipation component is expected to be that associated
with the out-of-plane angles of the edge defects. These angles
influence only a small region of magnetization immediately around
the cores of the half-antivortices, and so will not have large
contributions to the integral in Eq.~(\ref{eq-ccdef2}). Therefore,
we ignore $\Gamma_\theta$ in what follows.

\subsubsection{Connection with the two-mode approximation}

This model can be related back to the model in which only the vortex
is free to move by letting $k_2\rightarrow \infty$ and
$k_\theta\rightarrow \infty$. This will require $X_++X_-=X+\chi Y$,
$-\chi(X_+-X_-)=w$, and $\theta_+=\theta_-=0$ for all time, leading
to the reduced Lagrangian and Rayleigh function for just the vortex
position:
\begin{subequations}
\label{eq-Lv}
\begin{eqnarray}
L_v&=&-pG\dot{X}Y-U(X,Y),\\
R_v&=&\Gamma_X\dot{X}^2/2 + \Gamma_Y\dot{Y}^2/2
+ \Gamma_e\left(\dot{X}+\chi \dot{Y}\right)^2.
\end{eqnarray}
\end{subequations}
Identifying $\Gamma_{X}=\Gamma_{XX}-\chi\Gamma_{XY}$,
$\Gamma_{Y}=\Gamma_{YY}-\chi\Gamma_{XY}$, and
$\Gamma_{e}=\chi\Gamma_{XY}/2$, these are exactly the functions used
when only the components of the vortex position are soft modes
(Sec.~\ref{sec-wallmo}).  It becomes clear then that the off-diagonal
elements of the viscosity tensor in the 2-mode model arise directly
from the asymmetric equilibrium positions of the edge defects. By
comparing the identifications above with the viscosity values in
Table~\ref{table-Gamma} and Appendix~\ref{app-models}, one can see
that $\Gamma_e$ is indeed the dominant viscosity term, especially as
the width increases.
        \begin{figure}[tb]
        \includegraphics[width=\columnwidth]{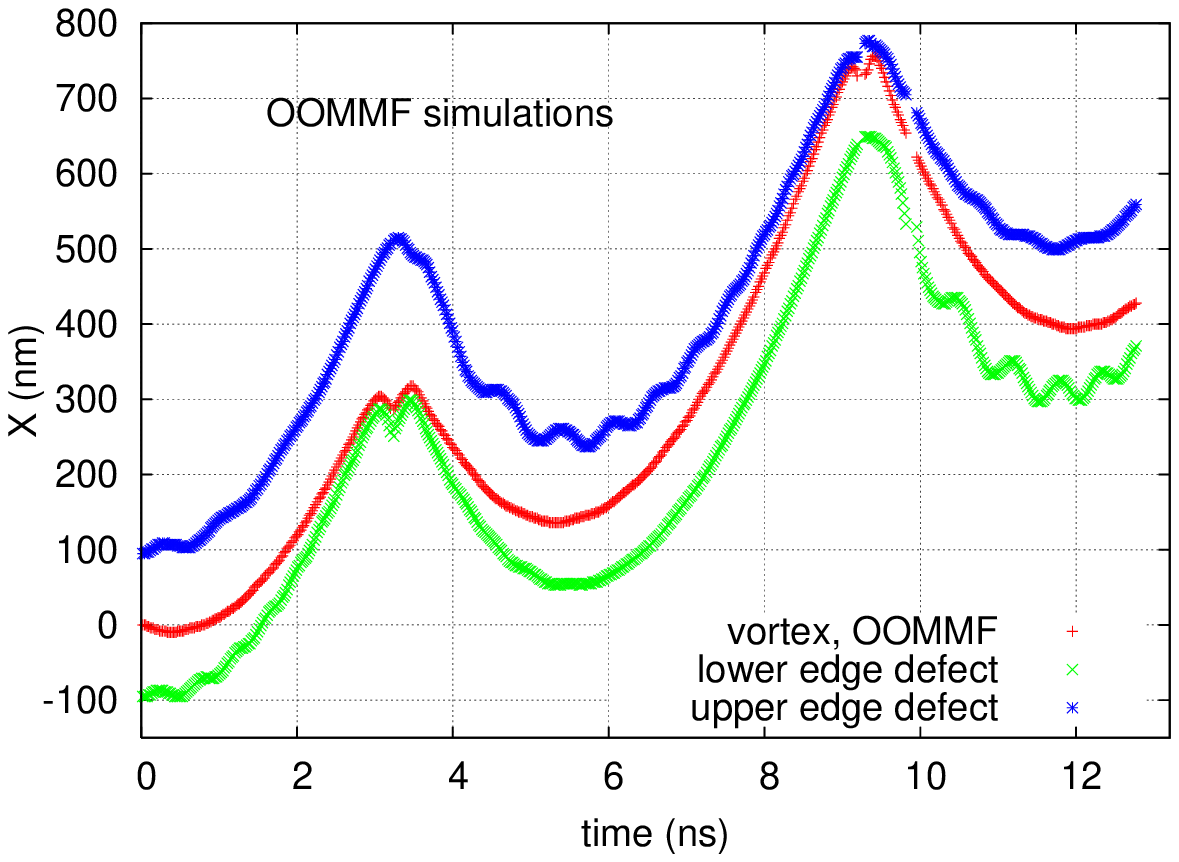}
        \includegraphics[width=\columnwidth]{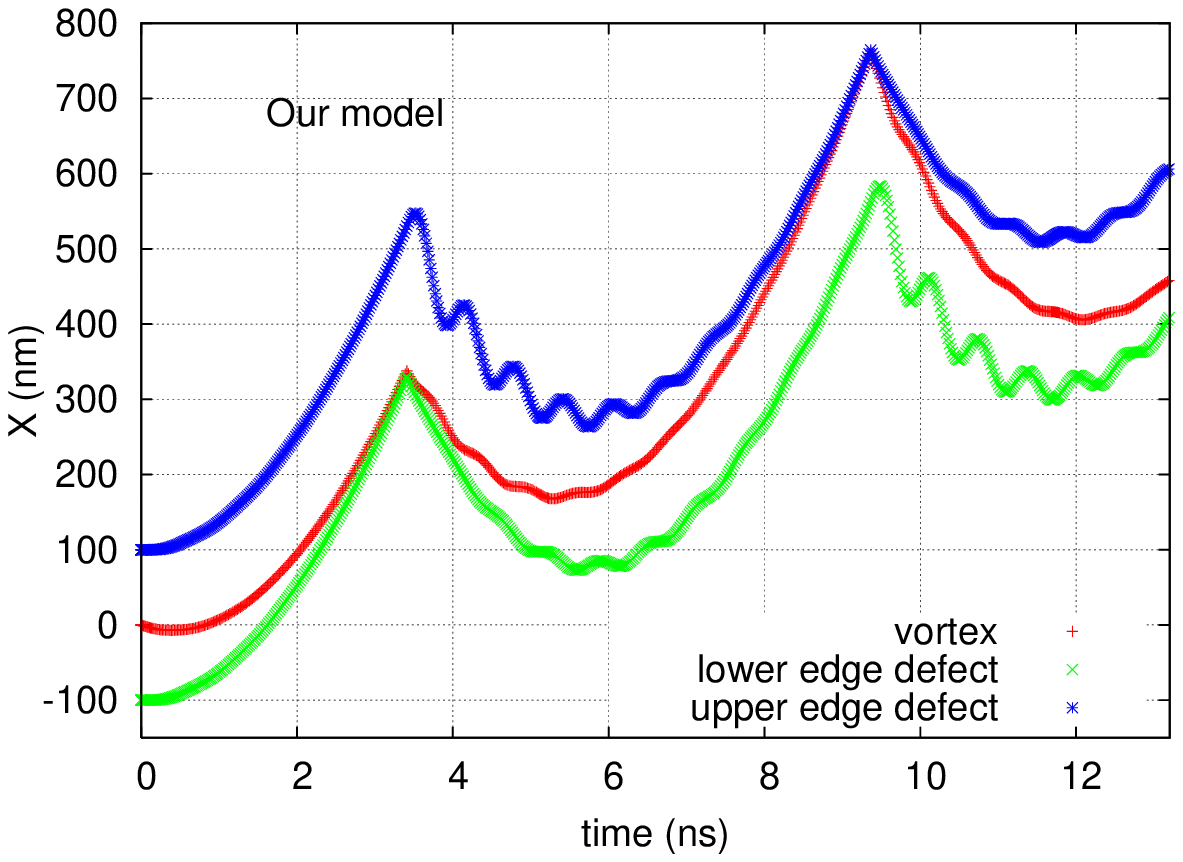}
        \caption{(Color online) Top: Numerically determined
            $x$-positions of the vortex and the two
            half-antivortices in a 20 nm thick, 200 nm wide
            permalloy strip as functions of time for $H=32$ Oe.
            Bottom: The longitudinal positions of the vortex and the
            two half-antivortices in the model with massive edge
            defects.  The vortex merges with an edge defect each
            time it hits the edge.  Note that the oscillation of
            that edge defect is suppressed compared to the other.}
        \label{x_t_reset}
        \end{figure}
Now that we have determined the viscosity values for the new model in
terms of the old one, we can check the descriptive value of the model
by testing the conservation of the total momentum in the numerical
data. This momentum is derived in the next section.

\subsubsection{Mass and momentum of the edge defects}

If we care only about the $x$-positions of the edge defects, we may
integrate out the values of $\theta_\pm$ as in Sec.~\ref{sec-1Deff} to
produce masses $m=G_2^2/k_\theta$ for the edge defects.  The
parameters $k_\theta$ and $G_2$ then appear only in this combination.
In effect, these masses occur because the moving wall is able to store
kinetic energy in the out-of-plane angle of the edge defects. The
Lagrangian then becomes:
\begin{equation}
\label{eq-U4m}
L = -pG\dot{X}Y -U(X,Y) +\sum_{b=\pm} (m\dot{X}_b^2/2-U_{e b}).
\end{equation}
The total $x$-direction momentum is then equal to
$$
P_x=-pGY+m(\dot{X}_++\dot{X}_-)
$$
It is influenced by two forces in the $x$-direction. First, viscous
drag acts on each of the defects, totaling to
        \begin{equation}
            F_d=-\Gamma_X\dot{X}-\Gamma_e\dot{X}_+-\Gamma_e\dot{X}_-.
        \end{equation}
Second, the external field applies a constant force $F_H=QH$ on the
wall.

As in the two coordinate model, the translational invariance of the
system implies that the internal forces between the defects will
have no influence over the total $x$ momentum. Note in particular
that the analysis here is independent of the spring constants $k$
and $k_2$. Thus
        \begin{equation}
            \frac{d P_x}{d t}=F_d+F_H,
        \end{equation}
This equation may be integrated to yield
        \begin{equation}
            P_x+\Gamma_{X}X+\Gamma_e\left(X_++X_-\right)-QHt=
            \mathrm{const}.
            \label{eq-xmom4}
        \end{equation}
We may determine the mass of the edge defects by fitting the left-hand
side of Eq.~(\ref{eq-xmom4}) to a constant.  The left-hand side of
this equation divided by the gyrotropic constant $G$ is plotted in the
bottom row of Fig.~\ref{fig-edgemass}. The figure shows that
Eq.~(\ref{eq-xmom4}) with $m=1.1\times10^{-23}$kg is satisfied by the
numerical simulation at \mbox{32 Oe} to nearly the accuracy of our
simulation, which had a 2 nm cell size.

\subsubsection{Oscillations of the wall width}

Insight into the width oscillations can be gained immediately from
the form of $L$ and $R$ in Eqs.~(\ref{eq-Lex}) and (\ref{eq-R}).
Defining the wall width $d$ by the difference $-\chi(X_+-X_-)$
between the forward and rear edge defects reveals that the conjugate
pair $(d,\bar{\theta})$, where $\bar{\theta}=(\theta_++\theta_-)/2$,
decouples from the remainder of the motion. Any deviation in $d$
from $w$ will decay over time with a time constant
$\tau_d=2m/\Gamma_e$, while oscillating due to the gyrotropic
coupling with the average out-of-plane angle of the defects.
        \begin{figure}[htbp]
        \includegraphics[width=\columnwidth]{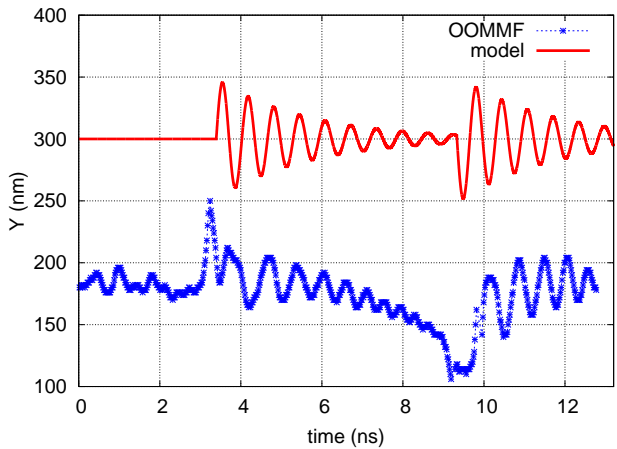}
        \caption{(Color online) Oscillations in the width of the
                wall in a strip with thickness $t = 20$ nm and width
                $w = 200$ nm in a field of $H = 32$ Oe. The upper
                curve shows the prediction of the model using the
                same parameters as in Fig.~\ref{x_t_reset}. It is
                offset vertically by 100 nm from the numerically
                determined result for clarity. }
        \label{fig-width}
        \end{figure}

        The equations of motion associated with the system described
        above can be solved to find the trajectories of the three
        defects. In order to take the collisions of the vortex with
        the edges of the strip into account, the polarization of the
        vortex is flipped every time it hits the edge. The effective
        width $w_{\rm{eff}} = w-2R_0$, with $R_0 = 10$ nm, is used
        to account for the short-range exchange interaction of the
        vortex with the edge. The polarization flip changes the sign
        of the momentum of the vortex, as in Sec.~\ref{sec-2mass}.
        In addition, when the vortex hits an edge, it actually must
        collide with one of the edge defects to be absorbed. The
        vortex is reemitted from that defect with the opposite
        polarization. As a result, the polarization of the edge
        defect changes as well. This causes oscillations in the wall
        width $d$ not because $d$ itself has changed, but because
        its conjugate momentum
        $\bar{\theta}$ has been altered.

        The solution shown in Fig.~\ref{x_t_reset} and
        Fig.~\ref{fig-width} uses a crude method to demonstrate this
        effect, simply reversing the value of the out-of-plane angle
        of an edge defect whenever a vortex hits it. Thus the vortex
        and the edge defect that it hits both have their momentum change
        sign during the collision. This is similar to the effect
        discussed in Sec.~\ref{sec-2mass}, in that it causes a
        sudden transfer of energy from the main mode, associated
        with the vortex motion, to ancillary modes associated with
        the oscillation of the edge defects. This effect is seen
         in Figs.~\ref{fig-width-osc} and \ref{fig-width},
        as the wall oscillation is
        reactivated every time the vortex hits the edge of the
        strip.

        The
        $x$ coordinates of the three topological defects over time are
        plotted in Fig.~\ref{x_t_reset} for the numerical solution
        and for the model in this section. The parameters used for
        the model are: $G^2/(k\Gamma_{XX})=8.75$ ns as measured in
        Fig.~\ref{fig-ynum}, $m=1.1\times10^{-23}$ kg as determined
        in Fig.~\ref{fig-edgemass}, $r=1.1$ to match the points
        where the vortex hits the wall, and $k_2/k=1.3$ to match the
        frequency of oscillations.

        The oscillations in the positions of the edge defects in the model
        and numerical solutions are comparable. In particular, the
        model correctly captures the fact that the oscillations in
        the farther edge defect from the vortex are larger in
        amplitude than those of the nearer one. This comes from the
        combination of flipping the sign of the vortex momentum with
        the resetting of the angle of the impacted edge defect. The
        oscillations caused by each effect interfere with one
        another. This interference is destructive for the near
        defect and constructive for the far one.

        Figure~\ref{fig-width} shows the oscillations in the width
        of the wall. As expected, these oscillations have the
        largest amplitude immediately after the vortex is emitted
        from one of the walls. The time constant of the decay is of
        the same order in both the model and numerical solutions.
        This agreement for the decay constant $\tau_d=2m/\Gamma_e$
        is expected, since this constant depends only on the edge
        defect mass measured in Fig.~\ref{fig-edgemass} and the
        viscosity coefficient $\Gamma_e$, which is known from the
        two-coordinate model.

        However, Fig.~\ref{fig-width} also reveals the limitations
        of the model described in this section.  While this model is
        useful in explaining the oscillations of edge defects, and
        the choice of coordinates shows excellent agreement with the
        $x$-momentum equation (\ref{eq-xmom4}), as shown in
        Fig~\ref{fig-edgemass}, there is still a significant
        discrepancy between the the theoretical prediction and the
        numerical simulation in Fig.~\ref{fig-width}.  The average
        wall width seen in numerical simulations changes as the
        vortex crosses the strip, an effect that is not predicted by
        the theory thus far. To account for this effect, additional
        terms must be added to Lagrangian~(\ref{eq-Lex}). In
        particular, this model thus far has ignored field-dependent
        contributions to the interactions between the defects, as
        well as terms of higher than quadratic order in the
        coordinates.

\section{Discussion}

We have explored the dynamics of a vortex domain wall in a magnetic
strip of submicron width.  We have applied the method of collective
coordinates\cite{tretiakov:127204} to the case when the wall has two
soft modes related to the motion of the vortex core. A simplified
model of the vortex domain wall described in this paper yields
solvable equations of motion.  The calculated mobility of the wall in
the steady-state viscous regime at low fields agrees well with the
value measured by Beach \textit{et al.}\cite{Beach05} The steady
motion breaks down when the equilibrium position of the vortex moves
beyond the edge of the strip.  The critical velocity depends strongly
on the magnetization length and the sample thickness, and weakly on
the width; its calculated value agrees reasonably well with the data
of Beach \textit{et al.}\cite{Beach05} The dynamics above the
breakdown changes its character from overdamped to underdamped. In
this regime the velocity sharply declines at first but later starts to
rise again as the field strength increases. The predicted high-field
mobility is reduced in comparison with the low-field value by the
factor $\mu_\mathrm{HF}/\mu_\mathrm{LF}\approx 0.01$; the
experimentally observed reduction is not as strong:
$\mu_\mathrm{HF}/\mu_\mathrm{LF} \approx 0.1$.\cite{Beach05}

We have compared the results of our theory to numerical simulations
of a moving vortex wall in a permalloy strip of width $w = 200$ nm
and thickness $t = 20$ nm.  We have found that the results obtained
via the collective-coordinate method are in quantitative
agreement with the simulations both in the viscous regime at low
applied fields and in the oscillatory regime at moderately high fields.
As the field strength increases further,
the velocity observed in numerical simulations diverges from the
predictions of the two-coordinate model.  We have traced the breakdown
of the two-mode theory to the appearance of four additional soft modes
associated with the coordinates and out-of-plane angles of the
edge defects (antihalfvortices).

As expected, the dynamics of a domain wall reduces to the motion of
the elementary topological defects that make up the wall---bulk
vortices and halfvortices at the edge.  Of particular importance is
the special kinematics of vortices caused by their nonzero skyrmion
numbers ($\pm 1/2$). On the formal level, the two coordinates of a
vortex core form a canonically conjugated pair.  As a consequence,
when the vortex is constrained in the transverse direction by a
parabolic potential, its longitudinal motion acquires inertia.  By
our estimates, in a strip of width $w = 200$ nm and thickness $t =
20$ nm a vortex wall has an effective mass of approximately
$10^{-22}$ kg. The mass of an edge defect is approximately
$10^{-23}$ kg. Saitoh \textit{et al.}\cite{Saitoh04} determined the
mass of a transverse wall in 70nm wide, 45nm thick permalloy to be
$6.6\times10^{-23}$ kg.

In our analysis, we have made some simplifying assumptions that
require further investigation. First, we have assumed that any vortex
absorbed by the edge is immediately reemitted. While this holds true
for our simulated strips of $w=200$ and $t=20$ nm, it is not always
the case. There may be short delays between absorption and reemission
during which the wall motion is again viscous; the higher mobility
during this period would tend to increase the drift velocity. This
type of motion has been observed in simulations of thinner strips by
Lee \textit{et al.}\cite{JYLee07}

Second, while we have described the onset of wall width oscillations
in Sec.~\ref{sec-highmo} as a typical new mode, decaying
oscillations are not the only type of new mode that can occur. In
particular, the number and dynamic characteristics of soft modes may
change discontinuously as additional vortices or antivortices are
created in the bulk of the strip.\cite{guslienko:2008prl} We have
observed numerically the creation and subsequent annihilation of a
vortex-antivortex pair adjacent to the original vortex of the wall.
As in the process described by Van Waeyenberge \textit{et
al.,}\cite{Stoll06} the pair creation mediates the flipping of the
polarization of the vortex and results in the reversal of the
gyrotropic force.  Once the pair is created, the antivortex moves
closer to the original vortex and together they form a topologically
non-trivial bound state (a skyrmion) of the type described in detail
by Komineas,\cite{komineas:2007} eventually decaying in a spin wave
explosion via a quasi-continuous process in which the conservation
of topological charge is violated.\cite{hertel:177202, Tretiakov07}
The new vortex, which has polarization opposite that of the
original, takes over as the central vortex of the domain wall. This
type of behavior may dominate the motion in very high fields,
causing the periodicity of the motion to deviate from the prediction
of Eq.~(\ref{eq-crosstime}).

\acknowledgments

The authors thank G.~S.~D. Beach, C.-L.~Chien, S.~Komineas, and
A.~Kunz, and M.~Tsoi for useful discussions, and M.~O.~Robbins for
sharing computational resources.  This work was supported in part by
NSF Grant No. DMR-0520491, by the JHU Theoretical Interdisciplinary
Physics and Astronomy Center, and by the Dutch Science Foundation NWO/FOM.

\appendix
\section{Walker's solution for a $180^\circ$ Bloch domain wall}
\label{app:walker}

We apply the method of collective coordinates to derive
Walker's classic result\cite{Walker74} for the dynamics of a $180^\circ$
Bloch domain wall in a uniform magnetic field.  The easy axis is $z$ and the
magnetization varies along the $x$ direction, $\mathbf M = \mathbf M(x,t)$.
The energy density per unit area in the $yz$-plane is
\begin{eqnarray}\label{eq-U-W}
U &=&\int d x \big[-\mu_0 H M \cos\theta + K_0 \sin^2\theta
\cos^2\phi
\nonumber \\
&& - K\cos^2\theta + A\left(\theta'^2+\sin^2\theta \, \phi'^2\right)
\big],
\end{eqnarray}
where primes denote derivatives with respect to $x$, $\mathbf M =
M(\sin{\theta} \cos{\phi}, \sin{\theta} \sin{\phi}, \cos{\theta})$,
$H$ is an external magnetic field applied along the easy axis, $K$
is the easy-axis anisotropy, and $K_0 = \mu_0 M^2/2$ is the shape
anisotropy reflecting the energy of a magnetic field $H_x = -M_x$
generated inside the wall.

On the basis of exact results for the steady state and numerical
simulations beyond the steady state, Walker parameterized a domain
wall by three collective coordinates: the center of mass $X$, the
width of the wall $\Delta$, and the (uniform in space) azimuthal
angle $\phi$.  His Ansatz,
\begin{equation}
\cos\theta=\tanh\left(\frac{x-X(t)}{\Delta(t)}\right),
\quad
\phi=\phi(t),
\label{eq:ansatz}
\end{equation}
substituted into Landau-Lifshitz-Gilbert equations, yielded an
approximate solution that reproduced the numerical results
remarkably well.  In Walker's solution the center of the wall $X$
and the angle $\phi$ were treated as independent dynamical
variables, while the width of the wall $\Delta$ adjusted
adiabatically to their instantaneous values.  We will use the
general formalism of Sec.~\ref{sec-cc} to derive the equations of
motion for the three modes and to deduce that the wall width is the
hardest mode of the three with a parametrically short relaxation
time, thus justifying the applicability of the two-mode
approximation in the oscillatory regime where the period $T$ and the
relaxation times of the $X$, $\phi$ and $\Delta$ modes satisfy the
inequality
\begin{equation}
\tau_\Delta \ll T < \tau_\phi < \tau_X=\infty.
\label{eq:2-modes-Walker}
\end{equation}

The equations of motion (\ref{eq-cc}) require the calculation of generalized
forces $F_i$ and the matrices $\Gamma_{ij}$ and $G_{ij}$ using the Ansatz
for the shape of the domain wall (\ref{eq:ansatz}).  The antisymmetric
gyrotropic matrix $G_{ij}$ has only two nonzero components,
\begin{equation}
G_{\phi X} = -G_{X \phi} = \pm 2 J
\end{equation}
for a domain wall with the asymptotic behavior $\cos{\theta} = \pm
1$ at $x = -\infty$.  The viscosity matrix $\Gamma_{ij}$ is
diagonal, with
\begin{equation}
\Gamma_{XX}=\frac{2\alpha J}{\Delta},
\quad
\Gamma_{\phi\phi}=2\alpha J \Delta,
\quad
\Gamma_{\Delta\Delta}=\frac{\pi^2\alpha J}{6\Delta}.
\end{equation}
Lastly, the generalized forces are
\begin{subequations}
\begin{eqnarray}
  F_X&=&2\mu_0 M H, \\
  F_\phi&=&K_0 \Delta \sin{2\phi}, \\
  F_\Delta&=&2 (A/\Delta^2- K - K_0 \cos^2\phi).
\end{eqnarray}
\end{subequations}

The width of the wall adjusts to the equilibrium value
$\Delta(\phi) = \sqrt{A/(K + K_0\cos^2\phi)}$
on the time scale $\tau_\Delta$ determined by its viscosity
$\Gamma_{\Delta\Delta}$ and stiffness
$k_\Delta = -\partial F_\Delta/\partial\Delta = 4A/\Delta^3$:
\begin{equation}
\tau_\Delta = \frac{\Gamma_{\Delta\Delta}}{k_\Delta}
 = \frac{\alpha \pi^2 J}{24 (K + K_0 \cos^2{\phi})}.
    \label{eq:tau_Delta}
\end{equation}

On longer time scales, the dynamical variables are the wall position
$X$ and the angle $\phi$:
\begin{subequations}
\begin{eqnarray}
F_X - \Gamma_{XX} \dot X + G_{X\phi} \dot \phi = 0,
\\
F_\phi(\phi) - \Gamma_{\phi\phi} \dot \phi + G_{\phi X} \dot X = 0.
\end{eqnarray}
\end{subequations}
The wall position $X$ is a zero mode with an infinite relaxation
time. The terminal velocity is $\dot X = F_X/\Gamma_{XX} =
\gamma\mu_0 H \Delta/\alpha$.  The angle reaches its equilibrium
value, determined by the condition $F_\phi(\phi) + G_{\phi X} \dot X
= 0$, on the time scale $\tau_{\phi}$.  Unlike the relaxation time
of the wall width (\ref{eq:tau_Delta}), $\tau_{\phi}$ is not
determined by the angle viscosity $\Gamma_{\phi\phi}$ and stiffness,
$k_\phi = -\partial F_\phi /\partial \phi = -2K_0\Delta
\cos{2\phi}>0$, alone.  The gyrotropic coupling to $X$ considerably
softens this mode extending the relaxation time,
\begin{equation}
\tau_\phi
= \frac{G_{\phi X}^2 + \Gamma_{\phi\phi} \Gamma_{XX}}{k_\phi \Gamma_{XX}}
= -\frac{2J(1+\alpha^2)}{\alpha K_0 \cos{2\phi}}
> \frac{2J}{\alpha K_0}.
\end{equation}

The ratio of the relaxation times,
\begin{equation}
\frac{\tau_\Delta}{\tau_\phi}
 = -\frac{\alpha^2 \pi^2 K_0 \cos{2\phi}}{48(1+\alpha^2)(K + K_0 \cos^2{\phi})}
 < \frac{\alpha^2 \pi^2 K_0}{48 K},
\end{equation}
is very small, $4 \times 10^{-4}$, in yttrium iron garnet, where
$\alpha \approx 10^{-2}$ and $K_0/K = 21$.\cite{Walker74}  It means
that Walker's solution has a considerable range of fields $H$ for which
the period of oscillations $T$ satisfies inequality
(\ref{eq:2-modes-Walker}) and thus the system has two soft modes,
$X$ and $\phi$.

\section{The gyrotropic tensor for a bulk defect}\label{app-gyro}

The gyrotropic tensor is given by Eq.~(\ref{eq-ccdef}).  If one
describes the motion of a bulk topological defect, such as a vortex or
an antivortex with polarization $p$, it is convenient to choose the
$X$ and $Y$ coordinates of that defect as the collective
coordinates. This leads to a non-zero contribution to $\hat{G}$,
\begin{equation}\label{eq-G}
    G_{XY}=J\int dV\,\m\cdot\left(
    \frac{\partial \m}{\partial X}\times
    \frac{\partial \m}{\partial Y}\right).
\end{equation}
Note that the integrand here is closely related to the
skyrmion density\cite{Polyakov75}
\begin{equation}
\rho = \frac{1}{4\pi} \m \cdot ( \partial_x \m \times
    \partial_y \m) =
    \frac{1}{4\pi}\frac{\partial(\cos\theta,\phi)}{\partial(x,y)}.
\end{equation}
In fact, if one makes the change of variables in Eq.~(\ref{eq-G})
such that $\frac{\partial}{\partial X}\rightarrow-
\frac{\partial}{\partial x}$ and $\frac{\partial}{\partial
Y}\rightarrow-\frac{\partial}{\partial y}$, the integrand becomes
$4\pi\rho$.

This substitution is allowed only if the magnetization texture moves
rigidly with the defect core. While this is not true everywhere
within a domain wall, it is valid near the core itself, where the
rigidity is forced by circulation of the magnetization. On these
short distances, the exchange interaction dominates and forces the
magnetization to be out-of-plane, overcoming the shape anisotropy
that would keep it in-plane. Because the exchange interaction is
dominant, the shape of the core is mainly independent of its
position. Fortunately, the rigid region near the core is the only
region that contributes to the integral in Eq.~(\ref{eq-G}). Because
the integrand is a triple product, all three components of the
magnetization need to be non-zero for a point to contribute. This
only occurs near the core.

Thus we find that $G_{XY}=4\pi J\int \rho dV  = 2\pi J t n p$, where
$n$ is the in-plane ($\mathrm{O}(2)$) winding number, which is $1$
for a vortex and $-1$ for an antivortex, and $p=M_z/\vert
M_z\vert=\pm 1$ is the polarization of the core.

\section{Modeling the vortex wall}\label{app-models}

\begin{figure}
\includegraphics[width=0.95\columnwidth]{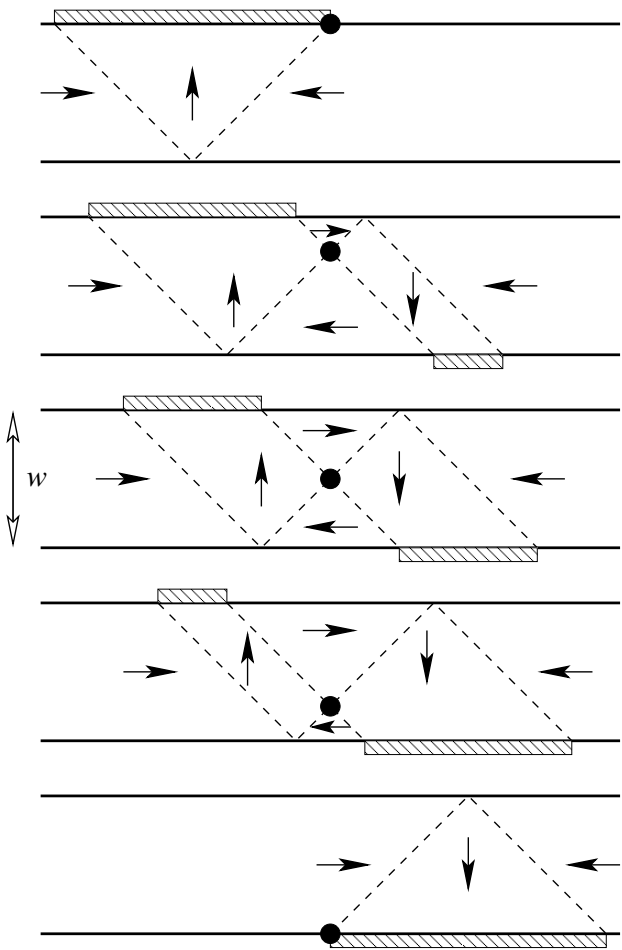}
\caption{A very simple (4-N\'eel-wall) model of the vortex domain
wall
 \cite{Chern-unpub} The vortex core is denoted by the filled
 circle. Four N\'eel walls separate regions of uniform
 magnetization. Shaded areas indicate a buildup of magnetic
 charge. The panels show states with a fixed longitudinal coordinate
 of the vortex $X = \mathrm{const}$; the transverse coordinate is $Y =
 w/2$, $w/4$, 0, $-w/4$, and $-w/2$ respectively.}
\label{fig-gw}
\end{figure}

In this appendix we describe two models for the vortex domain wall
based on two collective coordinates, the position of the vortex core
$(X,Y)$. The simpler model provides a pedagogical introduction and
the more sophisticated version is used to derive numerical values
for the parameters used in Sec.~\ref{sec-wallmo}.

In strips that support vortex domain walls, the dominant
contribution to the energy in the absence of an applied field is due
to magnetostatic interactions. For any domain wall in a strip of
width $w$, thickness $t$, and saturation magnetization $M$, there is
a total magnetic charge $Q=2\mu_0 M t w$ associated with the
wall.\cite{Youk06} In a vortex wall, nearly all of this charge is
expelled to the edge of the strip. We thus focus on models in which
there is no bound magnetic charge in the bulk of the strip.

We begin with a simple model in which a vortex wall consists of four
straight N\'eel walls separating regions of uniform magnetization
(Fig.~\ref{fig-gw}). Two of these walls cross at the vortex core,
and each edge defect has two walls emanating from it at $45^\circ$
angles from the edge. Despite its simplicity, the model captures
some of the essential features of the vortex wall, including the
chiral properties that lead to off-diagonal terms in the viscous
tensor and the transverse component of the Zeeman force on the
vortex.

\begin{figure}[htbp]
    \includegraphics[width=0.98\columnwidth]{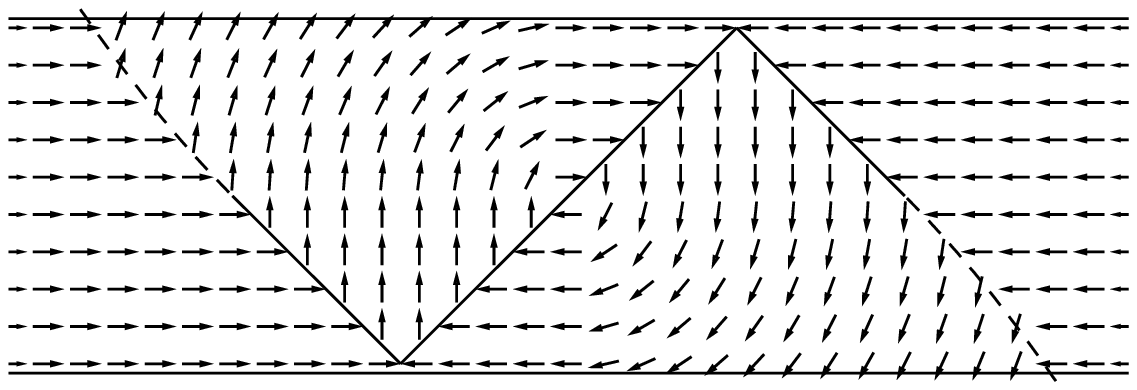}
    \includegraphics[width=0.98\columnwidth]{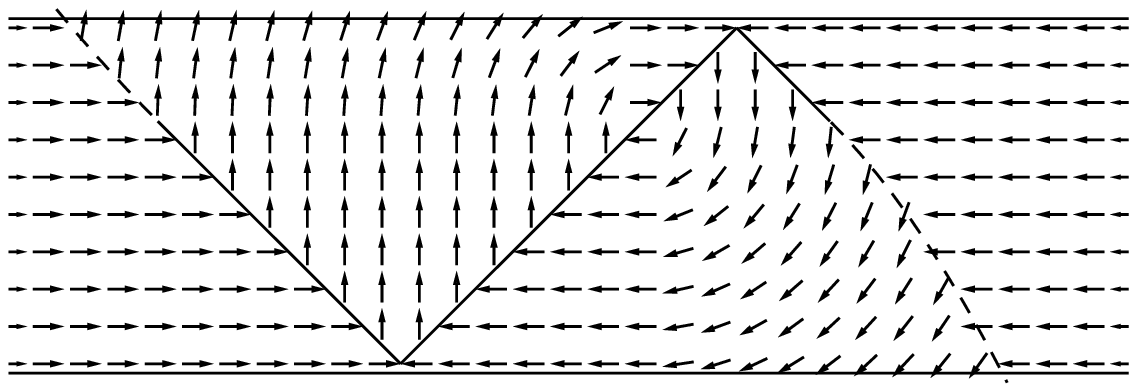}
    \includegraphics[width=0.98\columnwidth]{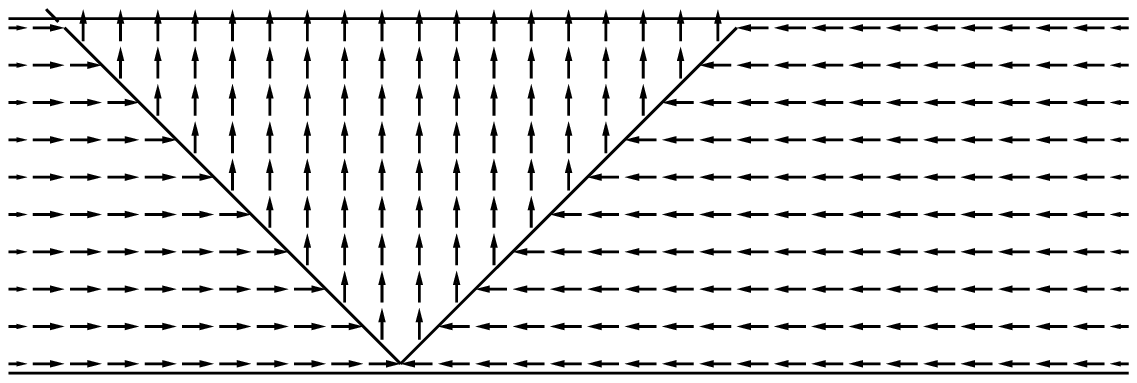}
    \caption{3-N\'eel-wall model of the vortex domain wall
        proposed in Ref.~\onlinecite{Youk06}.  Solid lines denote
        straight N\'eel wall regions and dashed lines denote
        regions in which the N\'eel walls are parabolic.}
    \label{fig-youk}
\end{figure}

While the 4-N\'eel-wall model described above correctly determines
the form of the free energy and the sign of the off-diagonal
viscosity term $\Gamma_{XY}$, it significantly overestimates the
magnitude of the viscosity coefficients. This is mainly due to the
inclusion of two N\'eel walls intersecting at the vortex core. In
the more realistic model first proposed by Youk \textit{et
al.}\cite{Youk06} one of these walls is replaced with two regions
where magnetization rotates gradually (Fig.~\ref{fig-youk}). In
these curling regions, the magnetization angle is given by
$\phi=\beta+\chi\pi/2$, where $\beta$ is the angular coordinate
around the vortex core, originating at the $x$-axis, and $\chi=\pm
1$ is the chirality of the vortex. As a result, we are left with
three N\'eel walls, with only one passing through the vortex. These
walls are straight where they separate two uniform regions and
parabolic where they separate a uniform and a curling region. This
prevents bound magnetic charge in the bulk of the system.

We proceed in this appendix to derive the components of the
viscosity tensor $\Gamma_{ij}$ and the values of the parameters $r$
and $k$ used in the energy (\ref{eq-U}). In each case we begin with
a brief explanation using the simpler model before deriving the
numerical values using the model of Youk \textit{et al}.

\subsection{Viscosity tensor}

    As long as the magnetization vector lies in the plane of the strip
    the definition of the viscosity tensor (\ref{eq-ccdef})
    may be expressed in terms of the azimuthal angle
    $\phi$ characterizing the magnetization:
    \begin{equation}
        \Gamma_{ij}=\alpha J t\int d^2 x \,
        \frac{\partial \phi}{\partial\xi_i}
        \frac{\partial \phi}{\partial\xi_j},
        \label{eq-Gamma-def}
    \end{equation}
    where $J = M/\gamma$ is the density of angular momentum and $t$ is
    the thickness of the film.

    The largest contribution to the viscosity comes from regions where
    the magnetization angle depends strongly on the collective
    coordinates $X$ and $Y$. Note that the translational symmetry of
    the problem means that $\phi(x,X,y,Y)=\phi(x-X,y,Y)$, allowing us
    to make the replacement $\partial \phi/\partial
    X=-\partial\phi/\partial x$ in all cases, regardless of the model.
    In the simple model (Fig.~\ref{fig-gw}) an infinitesimal
    shift in $X$ or $Y$ affects magnetization only in the vicinity of
    the N\'eel walls.

    We begin by considering the contribution to the viscous tensor
    of the N\'eel walls that emanate from the vortex core at $\pm
    45^\circ$. Near these walls, $\phi(x,y,X,Y) = f(x-X \mp y \pm Y)$.
    In these regions, derivatives with respect to collective coordinates
    may be reduced to ordinary gradients: $\partial \phi/\partial X =
    -\partial \phi/\partial x = -f'$ and $\partial \phi/\partial Y
    = -\partial \phi/\partial y = \pm f'$. As a result, the tensor
    components are equal to each other, up to a sign:
    \begin{equation}
        \Gamma_{XX}^N = \Gamma_{YY}^N = \mp \Gamma_{XY}^N
        = \alpha J t\int d^2x \, {f'}^2.
        \label{eq-f-prime}
    \end{equation}
    Opposite signs of the off-diagonal component
    $\Gamma_{XY}$ can be understood by noting that, as the vortex moves
    along $Y$, the two N\'eel walls that pass through it
    shift along $X$ in opposite directions, creating
    equal and opposite viscous forces in the $X$ direction.

As shown in Fig.~\ref{fig-gw}, the two peripheral N\'eel walls move in
the same direction as the central N\'eel wall perpendicular to
them. Since they are also of the same length as that wall, they have
the same contribution to the viscous tensor.  Adding the contributions
of all four N\'eel walls yields a total
    \begin{equation}
        \Gamma_{XX} = \Gamma_{YY} = 2 \chi \Gamma_{XY}, \label{eq-Gamma-rel}
    \end{equation}
independent of the vortex position (for the simple model).

To obtain the absolute values, note that the expressions for
viscosities (\ref{eq-f-prime}) coincide, up to a constant factor,
with the exchange energy of the N\'eel wall, which has been
calculated, e.g., in Ref. \onlinecite{Chern06}.

This energy may be determined by integrating the surface tension of
the N\'eel wall, given by \cite{Chern06}
\begin{equation}
\label{tension} \sigma = 2A\int dv \left(\frac{\partial
\phi}{\partial
        v}\right)^2=
\frac{2\sqrt{2}A}{\lambda}\left(\sin{\phi_0}-\phi_0\cos\phi_0\right),
\end{equation}
where $v$ is a coordinate perpendicular to the wall, and $2\phi_0$
is the angle of rotation across the wall. For a straight wall this
angle is constant along the entire wall and the remaining
integration along the wall is trivial. After an appropriate
normalization we obtain viscosity coefficients
    \begin{equation}
        \Gamma_{XX} = \Gamma_{YY} = 2\chi\Gamma_{XY}
        = 0.608 \alpha J tw/\lambda
            \label{eq-Gamma}
    \end{equation} for the simple model, where the exchange length
     $\lambda = \sqrt{A/\mu_0 M^2} = 3.8$ nm in permalloy.

    For a more accurate computation of the viscosity tensor, we turn to
    the model of Youk \textit{et al.}\cite{Youk06} (see
    Fig.~\ref{fig-youk}).  In this model, there are seven areas that
    contribute to the viscosity: 3 straight N\'eel wall segments, 2 parabolic
    N\'eel wall segments, and 2 regions of the bulk in which the magnetization
    curls.

    Eq.~(\ref{eq-f-prime}) still holds for the straight N\'eel wall
    regions. However, one of the N\'eel walls passing through the
    vortex is now absent, and the outer walls have half the length
    of the inner. Their viscosity contribution is equal to that of the inner
    wall. As a result, the total contribution from the
    straight segments of N\'eel wall is
    \begin{equation}
        \Gamma^S_{XX} = \Gamma^S_{YY} = \chi\Gamma^S_{XY} = 0.304 \alpha J t
            w/\lambda. \label{eq-Gamma-straight}
    \end{equation}

    In a similar fashion, we may calculate the viscosity of the
    parabolic N\'eel wall segments. There is a slight complication in this
    case as these segments deform as the vortex moves across the strip.

    The parabolic N\'eel walls in Fig.~\ref{fig-youk} are
    described by the equation $(x-X)^2=(2y\pm w)(2Y\pm w)$ for
    the upper and lower regions, respectively.\cite{Youk06} Note
    that when the vortex moves straight up, the distance that a
    given point on the parabolic N\'eel wall moves is dependent
    on its $y$ coordinate.  For an infinitesimal displacement
    $dY$ of the vortex upward from the center, a point on the
    parabolic N\'eel wall is moved to the left by $dY\sqrt{1 \pm
    2y/w}$. Thus, along these walls $\partial \phi/\partial Y=-
    (\partial \phi/\partial X) \sqrt{1\pm2y/w}$.

    In the case of the parabolic N\'eel walls, the angle $\phi_0$ changes
    along the length of the wall. For the lower wall,
    $\phi_0=(\arctan(y/\sqrt{w(w-2y)}))/2 +\pi/4$. This is equal to the
    angle of the wall normal away from the $x$-axis.  Therefore,
    \begin{equation}
        \int dx \left(\frac{\partial \phi}{\partial x}\right)^2
        = \frac{\sqrt{2}}{\lambda}(\sin{\phi_0}-\phi_0\cos\phi_0 )\cos\phi_0
    \end{equation}
    in the region near the lower parabolic N\'eel wall. We
    assume that $\phi_0$ varies slowly on the scale of the N\'eel
    wall width. The variation of $\phi_0$ along the wall leads to
    parametrically small, width independent corrections to the N\'eel
    wall viscosity calculated here.

    Making the usual replacement $\partial \phi/\partial X=-\partial
    \phi/\partial x$, we can perform the integration in
    Eq.~(\ref{eq-Gamma-def}) numerically for each tensor component.  When
    the vortex is centered on the strip, the contribution of the parabolic
    walls to the viscosity is given by
    \begin{subequations}
    \begin{eqnarray}
            \Gamma^P_{XX}&=& 0.057\alpha J t w/\lambda,\\
            \Gamma^P_{YY}&=& 0.083\alpha J t w/\lambda,\\
            \Gamma^P_{XY}&=& 0.069\alpha \chi J t w/\lambda.
    \end{eqnarray}
    \end{subequations}

    The final contribution to the viscosity comes from those bulk
    regions in which the magnetization curls. In these regions, it is
    most convenient to carry out the integration of
    Eq.~(\ref{eq-Gamma-def}) in cylindrical coordinates
    $x-X=R\cos\beta$, $y-Y=R\sin\beta$. When the vortex is centered on
    the strip, we find
    \begin{eqnarray}\label{cyl-Gamma}
        \Gamma_{ij} &=& \alpha J t\int^{\beta_0}_{-\pi/2}
            d\beta\int^{-w/(2\sin\beta)}_{r_0}dR\,
            \frac{\partial \phi}{\partial\xi_i} \frac{\partial
            \phi}{\partial\xi_j}
            \nonumber\\
            &&+ \alpha J t\int^0_{\beta_0} d\beta\int^{w/(1+\sin\beta)}_{r_0}dR
            \,\frac{\partial \phi}{\partial\xi_i} \frac{\partial \phi}{\partial\xi_j},
    \end{eqnarray}
    where $\tan\beta_0=-1/\sqrt{8}$ and $r_0$ is the radius of the vortex
    core.

    The magnetization in these regions is given simply by
    $\phi=\beta+\pi\chi/2$, so that $\partial \phi/\partial Y=
    -(\cos\beta)/R$, and $\partial \phi/\partial X= (\sin\beta)/R$.
    Integrating (\ref{cyl-Gamma}) numerically gives
\begin{subequations}
\begin{eqnarray}
        \Gamma^C_{XX}&=&\alpha J t
            \left(\frac{\pi}{4}\ln\frac{w}{r_0}-0.398\right),\\
        \Gamma^C_{YY}&=&\alpha J t
            \left(\frac{\pi}{4}\ln\frac{w}{r_0}-0.010\right),\\
        \Gamma^C_{XY}&=&-\alpha\chi J t
            \left(\frac{1}{2}\ln\frac{w}{r_0}-0.133\right).
\end{eqnarray}
\end{subequations}

The symmetry of the domain wall when the vortex is centered on the
strip implies that the contribution from both bulk curling regions is
the same. Summing up all the contributions, we find
\begin{subequations}
\label{eq-Gamma-youk}
\begin{eqnarray}
        \!\!\!\!\Gamma_{XX}&=&\alpha J t
        \left(0.418\frac{w}{\lambda} +\frac{\pi}{2}\ln\frac{w}{r_0}-0.797
        \right),\\
        \!\!\!\!\Gamma_{YY}&=&\alpha J t
        \left(0.470\frac{w}{\lambda} +\frac{\pi}{2}\ln\frac{w}{r_0}-0.020
        \right),\\
        \!\!\!\!\Gamma_{XY}&=&\alpha \chi J t
        \left(0.442\frac{w}{\lambda} -\ln\frac{w}{r_0}+0.265
        \right).
\end{eqnarray}
\end{subequations}

It is instructive to compute the ratio of the viscous and gyrotropic
forces $\Gamma_{ij}/G$ where $G = 2\pi Jt$ is the gyrotropic constant.
Taking the vortex core radius $r_0 = \lambda = 3.8$ nm, we obtain the
dimensionless ratios listed in Table~\ref{table-Gamma}. The small
value of Gilbert's damping in permalloy, $\alpha \approx 0.01$,
\cite{Freeman98} leads to the dominance of the gyrotropic force in
strips with submicron widths.  The smallness of $\Gamma_{ij}/G$ can be
exploited to organize an expansion in powers of this small parameter.

\subsection{Free energy}

In our simple model, magnetic charges form two lines of lengths $w-2Y$
and $w+2Y$ with constant charge density per unit length $\rho = \mu_0
Mt$.  As the vortex moves off-center, charge builds up on one edge of
the strip or the other, leading to an increase in magnetostatic
energy. The energy $E(Y) = E(0) + kY^2/2 + \mathcal O(Y^4)$ has a
minimum at $Y=0$. (Here we chose $Y=0$ to be in the middle of the
strip.) This leads to a force $-kY$ that acts to keep the vortex
centered on the strip.

As can be seen from Fig.~\ref{fig-gw}, transverse motion of the
vortex core changes the total magnetization $M_x$ of the strip and
thus affects its Zeeman energy. As the vortex core crosses from the
top to the bottom edge of the strip, the Zeeman energy decreases
linearly with the $y$-position of the core by a total of $4\mu_0 HM
t w^2$. The dependence of the energy of the magnetic configuration
on the vortex core position $(X,Y)$ is therefore given by
    \begin{equation}
        U(X,Y)=-QH(X+2\chi Y) + E(Y).
        \label{eq-U-x}
    \end{equation}

The same principles hold true in the more realistic 3-N\'eel-wall
model.  As the vortex moves off center, the buildup of charge on one
side of the strip causes a restoring force that pushes the vortex back
toward the middle of the strip. As in 4-N\'eel-wall model, the Zeeman
energy is given by $-QH(X+r\chi Y)$. Because the states of the system
when the vortex is at either edge are the same as they are in the
simplified model (see Figs.~\ref{fig-gw} and~\ref{fig-youk}), the
average value of $r$ as the vortex crosses the strip is again 2.

This is most likely an overestimate, as the states shown in
Figs.~\ref{fig-gw} and~\ref{fig-youk} when the vortex is at an edge
have an extended edge defect with a core radius equal to the width of
the strip. In reality, the half-vortex core is less extended due to
the exchange cost of the surrounding N\'eel walls. This smaller core
for the edge defect means that the total charge on the wall moves less
in the $X$ direction when the vortex crosses the strip than it does in
either of the models listed here. This leads to a smaller value for
$r$. By fitting the vortex trajectory in numerical simulations to that
of our collective coordinate analysis, we found $r\approx1.1$ in a
strip with the width $w = 200$ nm and thickness $t = 20$ nm.

In what follows, we derive the energy of the vortex domain wall in the
model proposed by Youk \textit{et al.}\cite{Youk06} to the second
order in the displacement of the vortex from the center of the strip.
As a result, we determine the value of the stiffness constant $k$.
There are four contributions to the energy in this model: the Zeeman
energy of the magnetization in the external field, the magnetostatic
energy of the bound charges on the edges of the strip, the exchange
energy of the bulk curling regions around the vortex, and the
integrated N\'eel wall tension of the three N\'eel walls emanating
from the topological defects.

\begin{figure}
\includegraphics[width=\columnwidth]{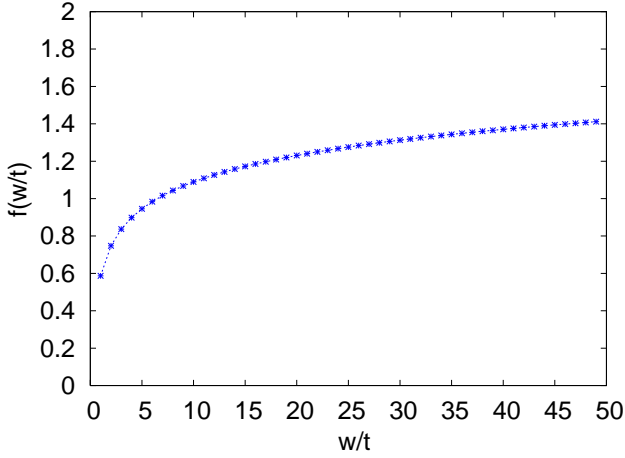}
\caption{Value of the scaling function $f$ for various values of the
        aspect ratio $w/t$.}
\label{fig-f}
\end{figure}

    If $(0,Y)$ is the position of the vortex, with $Y$ measured from
    the center of the strip, then the magnetization angles in the
    upper and lower curling regions are given by
    $\phi_\pm=\arctan((y-Y)/x)\mp\pi/2$. Thus there are charge
    densities $\rho_\pm(x)=\mu_0 M \sin\phi_\pm(\mp w/2,x)= \mu_0 M
    \abs{x}/\sqrt{x^2+(w/2\mp Y)^2}$ spread out along the edges. The
    charge distributions are bounded in the center by the $X$ position
    of the vortex and outside by the points $L_\pm=\mp\sqrt{2w(w\pm
    2Y)}$ at which the parabolic N\'eel wall segments hit the edges.
    The self-energies of these charged regions are given by
    \begin{equation}
        \frac{1}{8\pi\mu_0}
        \int_0^{t}\!\!d z\!\!
        \int_0^{t}\!\!d z'\!\!
        \int_0^{L_\pm(Y)}\!\!\!\!\!\!\!\!\!\!\!\!d x
        \int_0^{L_\pm(Y)}\!\!\!\!\!\!\!\!\!\!\!\!d x'
        \frac{\rho_\pm(x)\rho_\pm(x')}{\sqrt{(x-x')^2+(z-z')^2}}, \nonumber
    \end{equation}
    The interaction energy of the two charged segments is given by
    \begin{equation}
        \frac{1}{4\pi\mu_0}
        \int_0^{t}\!\!d z\!\!
        \int_0^{t}\!\!d z'\!\!
        \int_0^{L_+(Y)}\!\!\!\!\!\!\!\!\!\!\!\!d x
        \int_0^{L_-(Y)}\!\!\!\!\!\!\!\!\!\!\!\!d x'
        \frac{\rho_+(x)\rho_-(x')}{\sqrt{(x+x')^2+w^2+(z-z')^2}}.
        \nonumber
    \end{equation}
    Expanding to the second order in $Y$ and integrating numerically,
    we find that the magnetostatic energy
    $E_M=E_M(0)+\frac{k_M}{2}Y^2$. The value of $k_M$
    depends logarithmically on the aspect ratio $w/t$. For $w/t=10$,
    as in our numerical simulations,
    \begin{equation}
        k_M=1.08\frac{\mu_0M^2t^2}{w}.\nonumber
    \end{equation}
    In general,
    \begin{equation}
        k_M=f(w/t)\frac{\mu_0M^2t^2}{w},
    \end{equation}
    where the value of $f$ is plotted in Fig.~\ref{fig-f}.

In addition, there is a significant contribution to the stiffness $k$
from the N\'eel walls. When the vortex is at the edge, the N\'eel wall
energy is clearly lower than when it is at the center due to the
absence of the parabolic segments. Note that there is no change in the
energy of the straight N\'eel wall segments, because the total length
and tension of these segments stay the same as the vortex moves.

The parabolic N\'eel walls are described by the equation
$(x-X)^2=(2y\pm w)(2Y\pm w)$ for the upper and lower regions,
respectively. By symmetry, the total energy of the upper parabolic
segment, when the vortex position is $(X,Y)$, will be the same as that
of the lower parabolic segment when the vortex position is
$(X,-Y)$. Thus, the contribution of each parabolic segment to the
stiffness constant will be the same. We shall restrict our
consideration to the lower wall for simplicity.

It is convenient to derive the energy of the parabolic N\'eel wall
by switching to cylindrical coordinates around the vortex core.
Namely, $R=\sqrt{(x-X)^2+(y-Y)^2}$ and $\beta=\arctan((y-Y)/(x-X))$.
In these coordinates, the lower parabolic wall is given by
$R=(w-2Y)/(1+\sin\beta)$.

The surface tension of a N\'eel wall is given by Eq.~(\ref{tension}),
where $\lambda=\sqrt{A/\mu_0M^2}$.  In the case of the parabolic
N\'eel walls, the rotation angle $\phi_0$ is equal to the angle of the
wall normal away from the $x$-axis. This wall normal changes with
$\beta$. For the lower wall, $\phi_0=\beta/2+\pi/4$. The wall hits the
edge of the strip when $\beta=\beta_0(Y)=-\arcsin((w+2Y)/(3w-2Y))$.

    The line element along the parabolic segment is given by
    $dl=\sqrt{dr^2+r^2d\beta^2}=(w-2y)d\beta
    \csc\phi_0/(1+\sin\beta)$. Hence, the energy of the lower
    parabolic N\'eel wall is
    \begin{equation}
        E_{p-} = 2\sqrt{2}\mu_0M^2\lambda t(w-2Y)\int^0_{\beta_0(Y)}\!\!\!
        d\beta\,\frac{1-\phi_0(\beta)\cot\phi_0(\beta)}{1+\sin\beta}.
    \end{equation}
    Taking the second derivative of this energy with respect to $Y$ at $Y=0$
    yields the contribution
    \begin{equation}
        \frac{k_p}{2}=-\frac{4\mu_0M^2\lambda t}{3w}
    \end{equation}
    of the parabolic N\'eel walls to the spring constant $k$,
    where the factor of $2$ on the left hand side comes from the
    inclusion of the upper parabolic wall as well as the lower.

    Finally, there is a smaller contribution to the stiffness $k$ from
    the bulk regions of nonuniform magnetization. Again, the symmetry
    of the problem allows us to focus on the lower region and simply
    double our result to find the total contribution.  Using the
    cylindrical coordinates and definition of $\beta_0$, we find that
    the exchange energy of the lower region is given by
\begin{eqnarray}
E_{\mathrm{ex-}}&=&\mu_0M^2 t\int d^2x \,
\frac{\lambda^2}{r^2}\nonumber\\
&=&\mu_0M^2\lambda^2t
            \int^{\beta_0(Y)}_{-\pi/2}d\beta
            \ln \frac{-(w+2Y)}{2r_0 \sin\beta }
\nonumber\\
&&+ \mu_0M^2\lambda^2t
\int^0_{\beta_0}d\beta \ln \frac{(w-2Y)}{r_0 (1+\sin\beta)},
\nonumber
\end{eqnarray}
where $r_0$ is the radius of the vortex core.  If we again take the
second derivative with respect to $Y$ at $Y=0$, we obtain
    \begin{equation}
        \frac{k_{\mathrm{ex}}}{2}=
        -\left(\pi+\frac{4\sqrt{2}}{9}\right)
        \frac{2\mu_0M^2\lambda^2t}{w^2}.
    \end{equation}

In total, we find that
    \begin{eqnarray}\label{eq-k-compl}
        k&=& k_M + k_p +k_{\mathrm{ex}}
        \nonumber\\
        &\approx&\left[f(w/t)-\frac{8\lambda}{3t}
        -\frac{4\lambda^2}{tw}
        \left(\pi+\frac{4\sqrt{2}}{9}\right)\right]
        \frac{\mu_0M^2t^2}{w}.\nonumber\\
    \end{eqnarray}
    For $t=20$ nm, $\lambda=3.8$ nm, and $w=200$ nm, this gives
    \begin{equation}
        k=0.49\frac{\mu_0M^2t^2}{w}.
    \end{equation}

\bibliography{micromagnetics}
\end{document}